\documentclass[debug]{rmaa}

\usepackage[]{hyperref}
\hypersetup{colorlinks = true, linkcolor = blue, anchorcolor = red, citecolor = blue, filecolor = red, urlcolor = red, hypertexnames=true}
\usepackage{rotating}

\usepackage{paralist}

\usepackage{psfrag,color}

\usepackage[latin1]{inputenc}

\title{A study of the NGC 1193 and NGC 1798 open clusters using CCD UBV photometric and Gaia EDR3 data} 

\author{
  Talar Yontan,\altaffilmark{1}
  Hikmet \c{C}akmak,\altaffilmark{1}
  Sel\c{c}uk Bilir,\altaffilmark{1}
  Timothy Banks,\altaffilmark{2,3}
  Michel Ra{\'u}l,\altaffilmark{4}
  Remziye Canbay,\altaffilmark{5}
  Seliz Ko{\c c},\altaffilmark{5}
  Seval Ta{\c s}demir,\altaffilmark{5}
  H\"{u}lya Er\c{c}ay,\altaffilmark{5}
  Bur\c{c}in Tan{\i}k \"{O}zt\"{u}rk,\altaffilmark{5}
  \& Deniz Cennet Dursun\altaffilmark{5}}

\altaffiltext{1}{Department of Astronomy~\& Space Sciences, Istanbul University, Istanbul, Turkey.}

\altaffiltext{2}{Nielsen, Chicago, USA.}

\altaffiltext{3}{Harper College, Illinois, USA.}

\altaffiltext{4}{Observatorio Astronomico Naciona, Universidad Nacional Autonoma de Mexico, Ensenada, Mexico.}

\altaffiltext{5}{Institute of Graduate Studies in Science, Istanbul University, Istanbul, Turkey.}

\shortauthor{Yontan et al.}
\shorttitle{NGC 1193 and NGC 1798}

\fulladdresses{
    
    \item Talar Yontan, Hikmet \c{C}akmak \& Sel\c{c}uk Bilir: Istanbul University, Faculty of Science, Department of Astronomy and Space Sciences, 34119, Beyaz\i t, Istanbul, Turkey (talar.yontan@istanbul.edu.tr, hcakmak@istanbul.edu.tr, sbilir@istanbul.edu.tr).
    
    \item Timothy Banks: Nielsen, 200 W Jackson Blvd, Chicago, IL 60606, USA (tim.banks@nielsen.com); Department of Physical Science \& Engineering, Harper College, 1200 W Algonquin Rd, Palatine, IL 60067, USA (tbanks@harpercollege.edu).
    
    \item  Michel Ra{\'u}l: Observatorio Astronomico Naciona, Universidad Nacional Autonoma de Mexico, Ensenada, Mexico (rmm@astro.unam.mx).
    
    \item Remziye Canbay, Seliz Ko{\c c}, Seval Ta{\c s}demir, H\"{u}lya Er\c{c}ay, Bur\c{c}in Tan{\i}k \"{O}zt\"{u}rk, \& Deniz Cennet Dursun: Istanbul University, Institute of Graduate Studies in Science, Programme of Astronomy and Space Sciences, 34116, Beyaz{\i}t, Istanbul, Turkey (rmzycnby@gmail.com, seliskoc@gmail.com, tasdemir.seval@ogr.iu.edu.tr, hulyaercay5@gmail.com, burcin.tanik@istanbul.edu.tr, denizcdursun@gmail.com).
        }

\listofauthors{Talar Yontan, Hikmet \c{C}akmak, Sel\c{c}uk Bilir, Timothy Banks, Michel Ra{\'u}l, Remziye Canbay, Seliz Ko{\c c}, Seval Ta{\c s}demir, H\"{u}lya Er\c{c}ay, Bur\c{c}in Tan{\i}k \"{O}zt\"{u}rk, \& Deniz Cennet Dursun}

\indexauthor{Yontan, T.}
\indexauthor{\c{C}akmak, T.}
\indexauthor{Bilir, S.}
\indexauthor{Banks, T.}
\indexauthor{Ra{\'u}l, M.}
\indexauthor{Canbay, R.}
\indexauthor{Ko{\c c}, S.}
\indexauthor{Ta{\c s}demir, S.}
\indexauthor{Er\c{c}ay, H.}
\indexauthor{Tan{\i}k Ozt\"{u}rk, B.}
\indexauthor{Dursun, D.C.}

\abstract{We present photometric, astrometric, and kinematic studies of the old open star clusters NGC 1193 and NGC 1798. Both of the clusters are investigated by combining data sets from  {\it Gaia} Early Data Release 3 (EDR3) and CCD {\it UBV} observational data. Analysis of the radial distribution of stars through the cluster regions indicates that the cluster limit radii are $r_{\rm lim}=8'$ for both of the clusters. We determine the membership probabilities of stars considering {\it Gaia} EDR3 proper motion and trigonometric parallax data, resulting in 361 stars in NGC 1193 and 428 in NGC 1798 being identified as most likely cluster members, having membership probabilities greater than $P\geq 0.5$. Mean proper motion components are estimated as ($\mu_{\alpha}\cos \delta$, $\mu_{\delta}) = (-0.207\pm 0.009, -0.431\pm 0.008$) for NGC 1193 and ($\mu_{\alpha}\cos \delta$, $\mu_{\delta} )=( 0.793\pm 0.006, -0.373\pm 0.005$) mas yr$^{-1}$ for NGC 1798. $E(B-V)$ color excesses were derived for NGC 1193 as $0.150 \pm 0.037$ and for NGC 1798 as $0.505 \pm 0.100$ mag through the use of two-color diagrams. Photometric metallicities are also determined from two-color diagrams with the results of [Fe/H] = $-0.30 \pm 0.06$ dex for NGC 1193 and [Fe/H]=$-0.20\pm 0.07$ dex for NGC 1798. The isochrone fitting distance and age of NGC 1193 are $5562\pm 381$ pc and $4.6\pm 1$ Gyr, respectively. For NGC 1798, these parameters are $4451\pm 728$ pc and $1.3\pm 0.2$ Gyr. These ages indicate that NGC 1193 and NGC 1798 are old open clusters. The overall present day mass function slopes for main-sequence stars are found as $\Gamma=1.38\pm2.16$ for NGC 1193 and $\Gamma=1.30\pm0.21$ for NGC 1798, which are in fair agreement with the value of \citet{Salpeter55}. Kinematic and dynamic orbital calculations indicate that NGC 1193 and NGC 1798 belong to the thick-disk and thin-disk populations, respectively. In addition, both of the clusters were born outside the solar circle, and both orbit in the metal-poor region of the Galactic disk.}

\resumen{}

\addkeyword{Galaxy: open cluster and associations - 
Individual: NGC 1193, NGC 1798 - Galaxy: Disk stars - Hertzsprung-Russell (HR)}

\begin{document}
\maketitle
--------------------------------------------------------------------------------------------

\section{General}
\label{sec:intro}

Open clusters are identified as groupings of stars, beyond those found in a single multiple star system, that are bound together by their weak self-gravitational forces. As cluster stars are formed by the collapse of the same molecular cloud, their basic astrophysical parameters such as color excess, distance, metal abundance, and age are similar while their masses and luminosities can range widely. This paper concentrates on open star clusters inside our own galaxy's disk, which are often called `galactic clusters'. These properties make such open clusters important tools to investigate the structure, formation, and evolution of the Galactic disk as well as to give opportunities to enhance our understanding of stellar evolution models. In particular, the study of old open clusters can give insight into the kinematic properties and chemical structure of the Galactic disk \citep{Friel95}.

\subsection{NGC 1193}

In 1786 William Herschel discovered the open cluster NGC~1193 ($\alpha=03^{\rm h} \: 05^{\rm m} \: 56^{\rm s}\!\!.\,64$, $\delta=+44^{\circ} \: 22^{\rm '} \: 58^{\rm ''}\!\!.\,80,~l = 146^{\circ}\!\!.\,8143$, $b=-12^{\circ}\!\!.\,1624$), located in the constellation of Perseus \citep{Dreyer88}. Together with an angular size of $2'$, NGC 1193 has a dense central stellar concentration and is classified as II3m \citep{Ruprecht66}. \citet{King62} reported that NGC 1193 is likely to be old. In the study of \citet{Janes82}, NGC 1193 was identified as a dense, poorly studied open cluster with the angular diameter of $2'$. \citet{Kaluzny88} presented the first CCD {\it BV} photometric study of NGC 1193, identifying five possible blue struggler stars in the cluster. By {\it BV} isochrone fitting to the color magnitude diagram (CMD), they determined the color excess and distance to be $0.12 \leq E(B-V) \leq 0.23$ mag and $4.2 \leq d \leq 4.9$ kpc, respectively. Additionally the metallicity, distance module, and age of the cluster were adopted as $Z=0.01$, $(m-M)_{\rm V}=13.8$ mag, and $t=8 \times 10^{9}$ years. Through investigation of the cluster's color-magnitude diagram, \citet{Kaluzny88} indicated that the subgiant branch stars are more populous than red giant branch stars. Utilizing spectroscopic observations, \citet{Friel89} calculated the first radial velocity estimate for the cluster as $\langle V_{\rm r} \rangle = -82$ km s$^{\rm -1}$. \citet{Friel93} performed medium resolution spectroscopic analyses and estimated cluster metallicity as ${\rm [Fe/H]}=-0.50 \pm 0.18$ dex from four giant members. They also calculated radial velocities of these stars, whose values lie within the $-64 \leq V_{\rm r}\leq -103$ km s$^{\rm -1}$. \citet{Tadross05} used  photometric data of \citet{Kaluzny88} and astrometric data from the USNO-B1.0 catalog of \citet{Monet03} to determine the cluster's color excess $E(B-V)=0.10\pm0.06$, distance modulus $\mu = 13.90 \pm 0.10$ mag, distance $d = 5.25 \pm 0.24$ kpc, age $t = 8$ Gyr, and metallicity as $Z= 0.008$. Moreover, \citet{Tadross05} analysed the cluster with regard to the radial profile of \citet{Vanden60} and estimated the core radius as ${r}_{\rm c}=1'.4$ and the limiting radius as $r_{\rm lim}=6'.5$. \citet{Kyeong08} applied a fitting procedure of the theoretical isochrones of \citet{Bertelli94} to the color-magnitude diagrams based on CCD {\it UBVI} photometric data of NGC 1193. They calculated the color excess as $E(B-V)=0.19 \pm 0.04$ mag, metallicity ${\rm [Fe/H]}=-0.45 \pm0.12$ dex, true distance module $(m-M_{\rm V})_{\rm 0}=13.30 \pm 0.15$ mag, and the cluster age as $\log{t} (\rm yr)= 9.7 \pm 0.1$.

The {\it Gaia} mission \citep{Gaia16} has led to substantial improvements in the quality and precision of astrometric, photometric, and spectroscopic data. {\it Gaia} has provided precise astrometric, photometric, and spectroscopic data of nearly 1.8 billion stars. \citet{Cantat-Gaudin18} identified 215 most likely cluster members, using astrometric and photometric data of stars across locality of NGC 1193. In the study, they determined the mean proper motion of the cluster as ($\mu_{\alpha}\cos\delta$, $\mu_{\delta}$)=($-0.125\pm0.023, -0.329\pm0.019$) mas yr$^{\rm -1}$ and the trigonometric parallax as $\varpi = 0.159\pm0.009$ mas. \citet{Soubiran18} used the second {\it Gaia} data release \citep[{\it Gaia} DR2;][]{Gaia18} spectroscopy to identify a radial velocity measurement for one member star of NGC 1193, calculating its radial velocity as $\langle V_{\rm r} \rangle = -83.24\pm 0.51$ km s$^{\rm -1}$. In addition, \citet{Carrera19} determined the mean radial velocity of the cluster as $\langle V_{\rm r} \rangle = -85.16$ km s$^{\rm -1}$, based on APOGEE spectroscopic data for two member stars of the cluster. \citet{Donor20} analysed three cluster member stars using APOGEE DR16 spectroscopic data and calculated the radial velocity and metallicity of the NGC 1193 as $\langle V_{\rm r} \rangle = -84.7 \pm 0.2$ km s$^{\rm -1}$ km s$^{\rm -1}$ and ${\rm [Fe/H]}=-0.34 \pm0.01$ dex, respectively. Using {\it Gaia} DR2 data, they determined the mean proper motion components of the cluster as ($\mu_{\alpha}\cos\delta$, $\mu_{\delta}$)=($-0.22\pm0.10, -0.36\pm0.07$) mas yr$^{\rm -1}$.


\subsection{NGC 1798}

The open cluster NGC 1798 ($\alpha=05^{\rm h} 11^{\rm m} 39^{\rm s}\!\!.\,36$, $\delta=+47^{\circ} \, 41 ^{\rm '} 27^{\rm ''}\!\!.\,60,~l=160^{\circ}\!\!.\,7043$, $b=+04^{\circ}\!\!.\,8500$) was discovered in 1885 by Edward Barnard, located in the Auriga constellation \citep{Dreyer88}. With an angular size of about 5 arcmin, this cluster is classified as II2m with a central dense stellar concentration \citep{Ruprecht66}. Examination of the cluster's CMD reveals that the regions of the main sequence and red clump (RC) stars are more distinct than the red giant branch (RGB). Based on this morphological feature, \citet{Janes94} gave the age of NGC 1798 as 1.5 Gyr and distance as 3.44 kpc.

The first CCD {\it UBVI} photometric observations of the NGC 1798 were made by \citet{Park99}. The angular diameter of the cluster was given as 8.3 arcmin (10.2 pc), the color excess $E(B-V) = 0.51\pm0.04$ magnitude, distance $d = 4.2\pm0.3$ kpc, metallicity ${\rm [Fe/H]} = -0.47\pm 0.15$ dex, and age $t = 1.4\pm 0.2$ Gyr. \citet{Lata02} used the data of \citet{Park99} to determine the absolute magnitude and color indices for the $I$ band as $I(M_{\rm V}) = -4.86$, $I(U-V)_{\rm 0} = 0.97$,  $I(B-V)_{\rm 0}= 0.82$, and $I(V-I)_{\rm 0} = 1.14$ mag. \citet{Maciejewski07} obtained the structural and astrophysical parameters of 42 open clusters with CCD {\it BV} photometry. They determined the cluster's limiting radius $r_{\rm lim} = 9$ arcmin, the core radius $r_{\rm c} = 1.3\pm 0.1$ arcmin, central stellar density $f_{\rm 0} = 9.5\pm 0.28$ star arcmin$^{-2}$, and the background stellar density $f_{\rm bg} = 3.14\pm 0.05$ stars arcmin$^{-2}$. These researchers used the isochrones of \citet{Bertelli94}, obtaining the color excess of the cluster as $E(B-V)=0.37^{+0.10}_{-0.09}$, the distance modulus as $(m-M)=13.90_{-0.63}^{+0.26}$ mag, distance as $d=3.55_{-1.22}^{+0.64}$ kpc, and age being $\log t~(\rm yr)=9.2$. \citet{Ahumada07} examined 1,887 blue struggler star (BSS) candidates in 427 open clusters and identified 24 BSS in the direction of NGC 1798. They indicated that six of these BSS are massive and 18 are low mass stars. \citet{Carrera12} calculated the radial velocities of four open clusters including NGC 1798 by analyzing spectroscopic data of their member stars. By measuring Ca II lines, \citet{Carrera12} determined the mean radial velocity of NGC 1798 as $\langle V_{\rm r} \rangle = 2\pm10$ km~s$^{-1}$. They used six member stars in total, consisting of five RGB stars and one main sequence turn-off star. \citet{Oralhan15} analyzed CCD {\it UBVRI} photometric observations of 20 open clusters and obtained their astrophysical parameters. They determined the reddening, photometric metallicity, distance modulus, distance, and age of the NGC 1798 as $E(B-V)=0.47\pm0.07$ mag, ${\rm [Fe/H]} = -0.50\pm 0.28$ dex, $(m-M)_{\rm 0} = 12.70\pm 0.04$ mag, $d=3.47\pm 0.06$ kpc, and $t = 1.78\pm 0.22$ Gyr.

\citet{Cantat-Gaudin20} used photometric and astrometric data from the {\it Gaia} DR2 \citep{Gaia18} to determine astrometric and astrophysical parameters of 2,017 open clusters. They identified 218 member stars in NGC 1798. Considering these members they calculated mean proper-motion components and trigonometric parallaxes of the cluster as $( \mu_{\alpha}\cos\delta$, $\mu_{\delta} )= ( 0.913 \pm 0.011$, $-0.318 \pm 0.010$) mas yr$^{\rm -1}$ and $\varpi = 0.178 \pm 0.005$ mas. \citet{Liu19} used astrometric and photometric data of 78 member stars of NGC 1798 to calculate the mean proper-motion components, trigonometric parallaxes, and age of the cluster as ($\mu_{\alpha}\cos\delta$, $\mu_{\delta}$)= ($0.903 \pm 0.026$, $-0.400 \pm 0.295$) mas yr$^{\rm -1}$, $\varpi = 0.241\pm 0.026$ mas, and $t=1.7\pm 0.1$ Gyr. 

A number of studies explored open clusters using ground-based telescopes within the scope of spectroscopic survey programs \citep{Gilmore12, Conrad14, Maciejewski07, Kos18}. Within the context of the APOGEE survey, \citet{Donor18} utilized spectral observations of 259 cluster member stars in 19 open clusters including NGC 1798 and obtained the radial velocity and different metal abundance values of the stars. Analysing nine member stars in NGC 1798 \citet{Donor18} determined the mean radial velocity as $\langle V_{\rm r} \rangle =2\pm 1.7$ km~s$^{-1}$  and the iron abundance ${\rm [Fe/H]} = -0.18 \pm 0.02$ dex. \citet{Soubiran18} analysed {\it Gaia} DR2 spectroscopic data of four member stars in the cluster and obtained the mean radial velocity as $\langle V_{\rm r} \rangle =2.60\pm0.41$ km~s$^{-1}$. \citet{Donor20} analysed eight cluster member stars using APOGEE DR16 spectroscopic data and calculated the radial velocity and metallicity of the NGC 1798 as  $\langle V_{\rm r} \rangle = 2.7 \pm 0.8$ km s$^{\rm -1}$ km s$^{\rm -1}$ and ${\rm [Fe/H]}=-0.27 \pm0.03$ dex, respectively. Using {\it Gaia} DR2 data, they determined the mean proper motion components of the cluster as ($\mu_{\alpha}\cos\delta$, $\mu_{\delta}$)=($0.83\pm0.04, -0.31\pm0.04$) mas yr$^{\rm -1}$.

\begin{figure*}
\centering
\includegraphics[scale=0.265, angle=0]{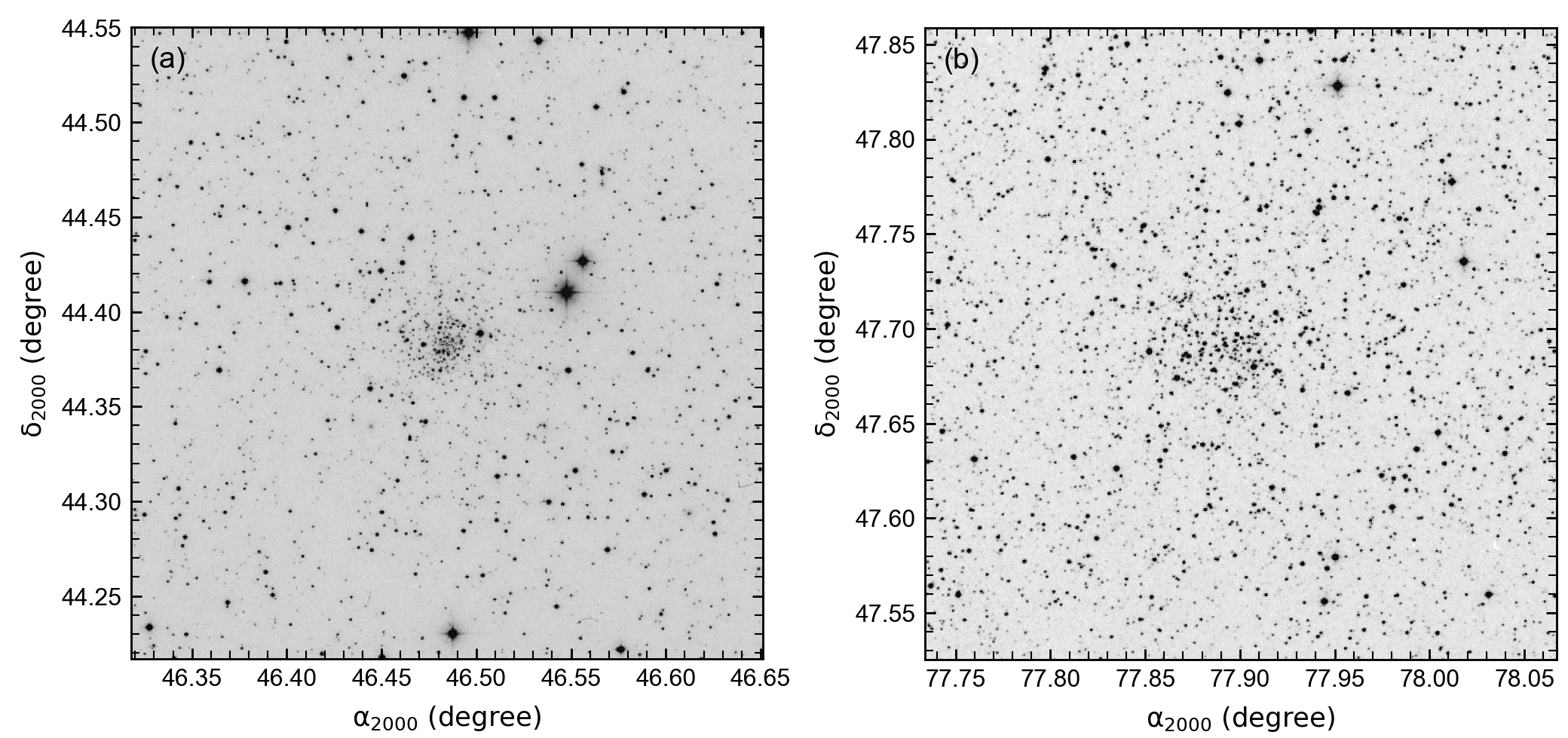}
\caption{Identification charts for NGC 1193 (left panel) and NGC 1798 (right panel), taken from the Leicester database and archive service (LEDAS).} 
\label{fig:ID_charts}
\end {figure*}


\section{Observations and data reductions}

\subsection{CCD {\it UBV} Photometric Data}
The observations of these two clusters, along with many others, were carried out at the San Pedro Martir Observatory,\footnote{https://www.astrossp.unam.mx/en/home/} as part of an ongoing {\it UBVRI} photometric survey of Galactic stellar clusters. The 84-cm ($f/15$) Ritchey-Chretien telescope was employed in combination with the Mexman filter wheel.

NGC 1193 was observed on 2013-09-19 with the ESOPO CCD detector (a $2048 \times 2048$ 13.5-$\mu$m square-pixels E2V CCD42-40 with a gain of $1.65 \: \mathrm{e^-}$/ADU and a readout noise of $3.8 \: \mathrm{e^-}$ at the $2 \times 2$ binning employed, providing an unvignetted field of view of $7.4 \times 9.3$ arcmin$^2$). Short and long exposures were taken to properly measure both the bright and faint stars of the fields. Exposure times were 2, 12, 120s for both {\it I} and {\it R}; 6, 30, 200 for {\it V}; 30, 100, 700s for {\it B}; and 60 and 1800s for {\it U}.

NGC 1798 was observed on 2009-11-01 with the SITE3 detector (a Photometrics $1024 \times 1024$ 24-$\mu$m square-pixels with a gain of $1.3 \:\mathrm{e^-}$/ADU and a readout noise of $6.8 \: \mathrm{e^-}$, giving an unvignetted field of view of $6.8 \times 6.8$ arcmin$^2$). Exposure times for {\it I} and {\it R} were 2, 12 and 120s in duration; 6, 30 and 200s for {\it V}; 30, 100 and 700s for {\it B}; and 60 and 1800s for {\it U}.

Landolt's standard stars \citep{Landolt09} were also observed in good sky conditions, at the meridian and at about two airmasses, to properly determine the atmospheric extinction coefficients. Flat fields were taken at the beginning and the end of each night and bias images were obtained between cluster observations. Data reduction with  point spread function (PSF) photometry was carried out by one of the authors (RM) with the IRAF/DAOPHOT packages \citep{Stetson87} and employing the transformation equations recommended, in their Appendix B, by \citet{Stetson19}.

\begin{figure*}
\centering
\includegraphics[scale=0.65, angle=0]{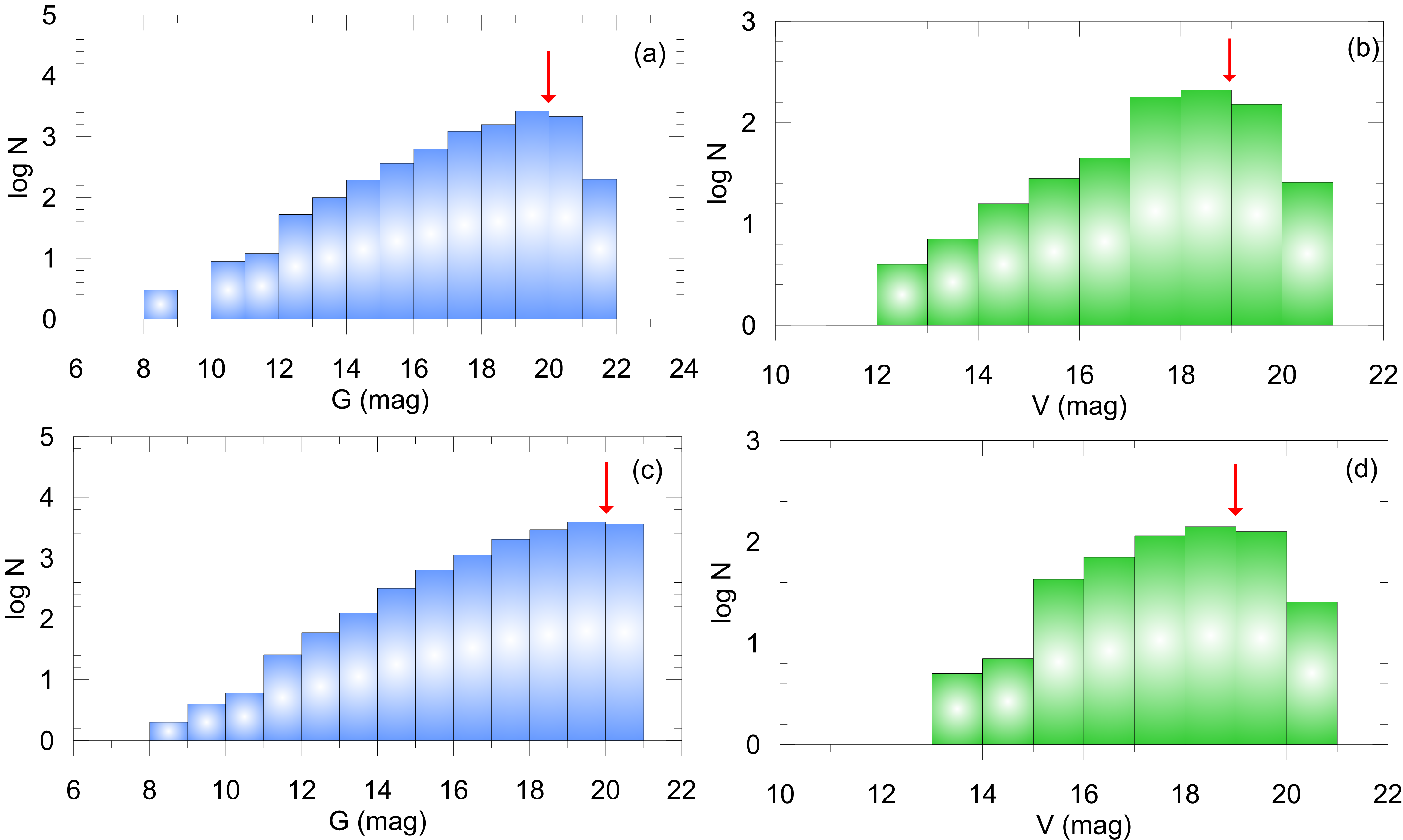}\\
\caption{Interval $G$ and $V$-band magnitude histograms of NGC 1193 (a, b) and NGC 1798 (c, d): The red arrows show the faint limiting apparent magnitudes in $G$ and $V$-bands.} 
\label{fig:histograms}
\end {figure*}


\section{Data Analysis}
\subsection{Gaia Astrometric and Photometric Data}

The (early) third data release of {\it Gaia} \citep[hereafter Gaia EDR3,][]{Gaia21} provides high quality astrometric and photometric data of nearly 1.5 billion celestial objects. Together with ground-based CCD {\it UBV} photometry we took into account {\it Gaia} EDR3 astrometric and photometric data to perform astrometric, photometric, and kinematic analyses of NGC 1193 and NGC 1798. We extracted such EDR3 data for all stars within regions of 20 arcmins about the centres of each cluster, using the coordinates given by \citet{Cantat-Gaudin20} ($\alpha=03^{\rm h} 05^{\rm m} 56^{\rm s}\!\!.\,64$, $\delta= +44^{\circ} 22^{\rm '} 58^{\rm''}\!\!.\,80$ for NGC 1193 and $\alpha=05^{\rm h} 11^{\rm m} 39^{\rm s}\!.\,36$, $\delta= +47^{\rm o} 41^{\rm '} 27^{\rm''}\!\!.\,60$ for NGC 1798). Thus we reached 9,141 stars within the magnitude range $7<G<23$ mag for NGC 1193 and 14,834 stars within $8<G<21$ mag for NGC 1798, respectively. 20 arcmin field of view optical images for the two clusters are presented in Figure~\ref{fig:ID_charts}. To construct photometric and astrometric catalogues for each cluster, we matched the {\it UBV} data to that from the {\it Gaia} EDR3 catalogue using stellar equatorial coordinates considering distances less than 5 arcsec. The mean difference in distances between the coordinates of stars in the matched catalogues was $\sim 0.08$ arc seconds for both clusters. Both resulting catalogues contain positions ($\alpha, \delta$), $UBV$ observational data (apparent $V$ magnitudes, color indices $U-B$, $B-V$), {\it Gaia} EDR3 astrometric ($\mu_{\alpha}\cos\delta, \mu_{\delta}$, $\varpi$) and photometric data ($G$, $G_{\rm BP}-G_{\rm RP}$), and membership probabilities ($P$) as calculated in this study (Table ~\ref{tab:all_cat}). Catalogues of CCD {\it UBV} photometric as well as {\it Gaia} photometric and astrometric data for all the detected stars in the cluster regions are available electronically for NGC 1193 and NGC 1798\footnote{The complete tables can be obtained from VizieR electronically}. Errors of the {\it UBV} and {\it Gaia} EDR3 photometric data were adopted as internal errors, being the uncertainties in the determination of the instrumental magnitudes of the stars. We calculated the  mean photometric errors separately as functions of $V$ and $G$ intervals, listing the results in Table~\ref{tab:photometric_errors} (on page \pageref{tab:photometric_errors}). It can be seen from the table that the mean internal {\it UBV} errors reach 0.08 mag for stars brighter than $V=20$ mag for both clusters. The mean internal errors of {\it Gaia} EDR3 photometry for stars brighter than $G=21$ mag reach 0.011 mag for NGC~1193 and 0.007 mag for NGC 1798.

\begin{sidewaystable}
\setlength{\tabcolsep}{4.2pt}
\renewcommand{\arraystretch}{1.2}
\scriptsize
  \centering
  \caption{\label{tab:input_parameters}
{The photometric and astrometric catalogues for NGC 1193 and NGC 1798}.}
    \begin{tabular}{cccccccccccc}
\hline
\multicolumn{11}{c}{NGC 1193}\\
\hline
ID	 & RA           &	DEC	        &      $V$	    &	$U-B$      & $B-V$	      &	$G$	          & 	$G_{\rm BP}-G_{\rm RP}$	 & 	$\mu_{\alpha}\cos\delta$ & 	$\mu_{\delta}$ & 	$\varpi$	& $P$ \\

	 & (hh:mm:ss.ss)           &	(dd:mm:ss.ss)	&      (mag)	    &	(mag)      & (mag)	      &	(mag)	          & 	(mag)	 & 	(mas yr$^{-1}$) & 	(mas yr$^{-1}$) & 	(mas)	&  \\
\hline
001 & 03:05:36.67 & +44:22:00.62 & 19.194(0.062) & ---          & 1.155(0.120) & 21.929(0.049) & 1.693(0.116) & ---           & ---           & ---            & ---  \\
002 & 03:05:37.44 & +44:22:01.67 & 18.984(0.044) & ---          & 0.893(0.083) & 18.777(0.003) & 1.080(0.043) & -0.587(0.247) & -0.251(0.223) & -0.083(0.241)  & 1.00 \\
003 & 03:05:38.33 & +44:19:33.12 & 19.162(0.056) & ---          & 1.475(0.140) & 18.254(0.003) & 2.116(0.035) & -4.640(0.181) & -0.566(0.165) & 1.009(0.177)   & 0.00 \\
004 & 03:05:38.40 & +44:21:46.02 & 19.286(0.067) & ---          & 1.262(0.137) & 18.592(0.003) & 1.759(0.040) & -4.072(0.220) & -2.553(0.200) & 0.640(0.214)   & 0.00 \\
005 & 03:05:38.74 & +44:24:25.10 & 18.631(0.034) & 0.194(0.170) & 0.993(0.065) & 18.418(0.003) & 0.969(0.032) & -0.426(0.220) & -0.410(0.235) & 0.185(0.220)   & 1.00 \\
... & ... & ...	& ... & ... & ... & ... & ... & ... & ... & ... & ...\\
565 & 03:06:18.54 & +44:18:51.83 & 17.758(0.017) & 0.458(0.088) & 0.881(0.026) & 17.510(0.003) & 1.137(0.014) &  1.640(0.128) & -2.394(0.097) & 0.433(0.119) & 0.00 \\
566 & 03:06:19.07 & +44:24:53.68 & 17.558(0.014) & 0.042(0.042) & 0.616(0.020) & 17.463(0.003) & 0.786(0.020) & -0.717(0.177) & -0.447(0.122) & 0.149(0.121) & 0.99 \\
567 & 03:06:19.29 & +44:19:29.22 & 15.451(0.006) & 0.058(0.011) & 0.697(0.008) & 15.287(0.003) & 0.908(0.005) & -5.602(0.039) & -2.523(0.030) & 0.604(0.036) & 0.00 \\
568 & 03:06:19.45 & +44:27:38.77 & 16.806(0.042) & 0.534(0.156) & 0.941(0.044) & 16.465(0.003) & 1.235(0.008) &  5.074(0.078) & -1.965(0.062) & 0.799(0.068) & 0.00 \\
569 & 03:06:19.77 & +44:18:25.54 & 19.881(0.101) & ---          & 0.658(0.152) & 19.655(0.005) & 1.061(0.069) &  0.254(0.530) &  0.166(0.412) & 0.264(0.521) & 0.88 \\
	
\hline
\multicolumn{11}{c}{NGC 1798}\\
\hline
ID	 & RA           &	DEC	        &      $V$	    &	$U-B$      & $B-V$	      &	$G$	          & 	$G_{\rm BP}-G_{\rm RP}$	 & 	$\mu_{\alpha}\cos\delta$ & 	$\mu_{\delta}$ & 	$\varpi$	& $P$ \\

	 & (hh:mm:ss.ss)           &	(dd:mm:ss.ss)	&      (mag)	    &	(mag)      & (mag)	      &	(mag)	          & 	(mag)	 & 	(mas yr$^{-1}$) & 	(mas yr$^{-1}$) & 	(mas)	&  \\
\hline
001 & 05:11:19.70 & +47:39:52.01 & 20.254(0.135) & ---           & 0.563(0.192) & 19.499(0.004) & 1.573(0.056) & 0.968(0.426) & -0.986(0.343) & 0.270(0.344) & 0.95 \\
002 & 05:11:19.73 & +47:38:35.08 & 19.765(0.089) & ---           & 1.182(0.171) & 19.278(0.004) & 1.575(0.048) & 0.919(0.372) & -0.611(0.283) &-0.051(0.289) & 0.91 \\
003 & 05:11:19.79 & +47:41:26.68 & 18.623(0.041) & ---           & 1.046(0.074) & 18.215(0.003) & 1.471(0.024) &-0.020(0.175) & -0.596(0.137) & 0.083(0.140) & 0.29 \\
004 & 05:11:20.00 & +47:43:15.08 & 20.046(0.151) & ---           & 0.920(0.208) & 19.369(0.004) & 1.604(0.057) & 1.623(0.331) & -2.917(0.270) & 0.418(0.279) & 0.00 \\
005 & 05:11:20.44 & +47:41:06.65 & 17.635(0.016) & ---           & 1.442(0.039) & 17.001(0.003) & 1.683(0.012) & 3.835(0.152) & -3.762(0.118) & 0.584(0.116) & 0.00 \\
... & ... & ...	& ... & ... & ... & ... & ... & ... & ... & ... & ...\\
486 & 05:11:58.43 & +47:39:28.67 & 16.202(0.024) &  0.423(0.045) & 0.494(0.032) & 16.057(0.003) & 0.747(0.006) & 0.920(0.051) & -0.354(0.042) & 0.229(0.042) & 1.00 \\
487 & 05:11:58.51 & +47:38:31.88 & 17.366(0.020) &  0.807(0.267) & 1.321(0.046) & 16.766(0.003) & 1.623(0.008) & 0.843(0.073) & -1.241(0.059) & 0.119(0.059) & 0.97 \\
488 & 05:11:58.73 & +47:38:58.99 & 17.628(0.024) &  0.286(0.162) & 0.913(0.040) & 17.158(0.003) & 1.322(0.008) & 0.805(0.098) & -3.354(0.077) & 0.354(0.074) & 0.00 \\
489 & 05:11:58.88 & +47:40:05.24 & 15.629(0.040) &  0.281(0.038) & 0.551(0.043) & 15.582(0.003) & 0.870(0.005) & 0.701(0.056) & -0.518(0.043) & 0.289(0.043) & 1.00 \\
490 & 05:11:59.15 & +47:42:42.85 & 20.344(0.191) & ---           & 0.931(0.298) & 19.339(0.004) & 1.461(0.057) & 0.414(0.358) &  0.057(0.294) & 0.906(0.305) & 0.18 \\
\hline
    \end{tabular}
      \label{tab:all_cat}%
\end{sidewaystable} 

\begin{table*}[!t]
\setlength{\tabcolsep}{3pt}
\renewcommand{\arraystretch}{0.8}
  \centering
  \caption{\label{tab:photometric_errors}
  {The mean internal photometric errors and number of measured stars in the corresponding $V$ apparent-magnitude interval for each cluster.}}
 \begin{tabular}{ccccc|cccc} 
      \hline \\
\multicolumn{5}{c}{NGC 1193} & \multicolumn{4}{c}{NGC 1798} \\ 
\hline 
\rule{0pt}{2.5ex} $V$ & $N$ & $\sigma_{\rm V}$ & $\sigma_{\rm U-B}$ & $\sigma_{\rm B-V}$ & $N$ & $\sigma_{\rm V}$ & $\sigma_{\rm U-B}$ & $\sigma_{\rm B-V}$ \\[0.5ex]
  \hline 
\rule{0pt}{2.5ex}
(8, 12]  & 1   & 0.004 & 0.007  & 0.007     & ---  & ---   & ---    & ---   \\
(12, 14] & 11  & 0.019 & 0.021  & 0.024     & 5    & 0.04  & 0.047  & 0.045 \\
(14, 15] & 16  & 0.007 & 0.012  & 0.010     & 7    & 0.022 & 0.055  & 0.035 \\
(15, 16] & 28  & 0.008 & 0.019  & 0.011     & 44   & 0.021 & 0.056  & 0.033 \\
(16, 17] & 45  & 0.012 & 0.037  & 0.016     & 70   & 0.020 & 0.070  & 0.033 \\
(17, 18] & 178 & 0.018 & 0.065  & 0.028     & 114  & 0.023 & 0.148  & 0.037 \\
(18, 19] & 209 & 0.034 & 0.120  & 0.057     & 140  & 0.045 & 0.230  & 0.075 \\
(19, 20] & 152 & 0.076 & 0.234  & 0.132     & 126  & 0.077 &  ---   & 0.141 \\
(20, 21] & 26  & 0.146 & ---    & 0.244     & 26   & 0.136 &  ---   & 0.242 \\ 
 \hline \\
\multicolumn{5}{c}{NGC 1193}  & \multicolumn{4}{c}{NGC 1798} \\
 \hline
\rule{0pt}{2.5ex}$G$ & $N$ & $\sigma_{\rm G}$ &  $\sigma_{G_{\rm BP}-G_{\rm RP}}$ &  &  $N$  & $\sigma_{\rm G}$ & $\sigma_{G_{\rm BP}-G_{\rm RP}}$  &\\[0.5ex]
 \hline
\rule{0pt}{2.5ex}
(5, 10]  & 5    & 0.003 & 0.006 &       & 6    & 0.003 & 0.005  &     \\
(10, 12] & 21   & 0.003 & 0.005 &       & 32   & 0.003 & 0.005  &     \\
(12, 13] & 52   & 0.003 & 0.005 &       & 59   & 0.003 & 0.005  &     \\
(13, 14] & 101  & 0.003 & 0.005 &       & 125  & 0.003 & 0.005  &     \\
(14, 15] & 196  & 0.003 & 0.006 &       & 315  & 0.003 & 0.005  &     \\
(15, 16] & 366  & 0.003 & 0.006 &       & 630  & 0.003 & 0.006  &     \\
(16, 17] & 634  & 0.003 & 0.010 &       & 1115 & 0.003 & 0.009  &     \\
(17, 18] & 1219 & 0.003 & 0.019 &       & 2026 & 0.003 & 0.017  &     \\
(18, 19] & 1588 & 0.004 & 0.044 &       & 2936 & 0.003 & 0.035  &     \\
(19, 20] & 2616 & 0.005 & 0.155 &       & 3943 & 0.004 & 0.075  &     \\
(20, 21] & 2144 & 0.011 & 0.232 &       & 3647 & 0.007 & 0.152  &     \\
(21, 23] & 198  & 0.027 & 0.378 &       & ---  & ---   & ---    &     \\
       \hline
  \end{tabular}%
 \label{tab:photometric_errors}%
\end{table*}%

To obtain precise astrophysical parameters, we identified photometric completeness limits for each cluster.  Stars fainter that these limits were not included in further analyses. $G$ and $V$ magnitude histograms were constructed to determine the photometric completeness limits for each clusters (see Fig.~\ref{fig:histograms}). Stellar counts decrease for magnitudes fainter than $G=20$ for both NGC 1193 (Fig.~\ref{fig:histograms}a) and NGC 1798 (Fig.~\ref{fig:histograms}c). Stellar counts reduce for magnitudes fainter than $V=19$ for NGC 1193 (Fig.~\ref{fig:histograms}b) and NGC 1798 (Fig.~\ref{fig:histograms}d), indicating that incompleteness (of stellar recovery) has set in. Thus, for both clusters, we adopted these values as the cluster photometric completeness limits.

\subsection{Structural Parameters of the Clusters}

\begin{figure}
\centering
\includegraphics[scale=0.335, angle=0]{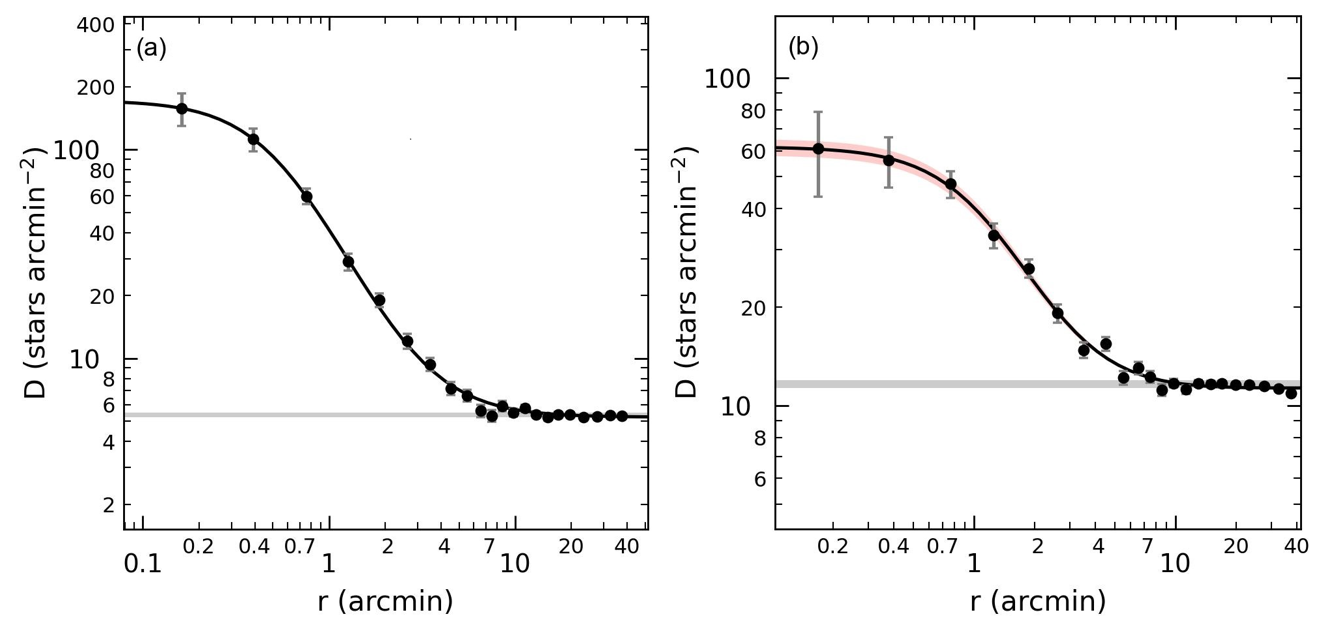}\\
\caption{Radial density profiles for NGC 1193 (a) and NGC 1798 (b). Errors were derived using the equation of $1/\sqrt N$, where $N$ represents the number of stars used in the density estimation. The solid lines represents the optimal \citet{King62} profiles. The background density level and its errors are the horizontal grey bands. The King fit uncertainty ($1\sigma$) is shown by the red shaded region.} 
\label{fig:king}
\end{figure}

We utilized Radial Density Profile (RDP) analysis to determine the structural parameters of the studied clusters. First, we specified many concentric rings out from the cluster center, using the central coordinates given by \citet{Cantat-Gaudin20}. Stellar densities ($\rho$) were estimated for each ring by dividing the number of stars within the photometric completeness limit ($G\leq 20$ mag) in it by the ring area. The resulting RDPs were fitted with \citet{King62} models via $\chi^2$ minimisation, giving estimates for the core, limiting, and effective radii of each cluster. The \citet{King62} model is described as $\rho(r)=f_{\rm bg}+[f_{\rm 0}/(1+(r/r_{\rm c})^2)] $ where $r$ is the radius from the cluster centre, $f_{\rm bg}$ the background density, $f_{\rm 0}$ the central density, and $r_{\rm c}$ the core radius. See Figure~\ref{fig:king} for each cluster's RDP together with the best fitting \citet{King62} model to it. As a result of the fitting procedure, we inferred central stellar density, core radius and background stellar density as $f_{\rm 0}=166.865\pm 1.573$ stars arcmin$^{-2}$, $r_{\rm c}=0.526\pm 0.009$ arcmin and $f_{\rm bg}=5.225\pm 0.124$ stars arcmin$^{-2}$ for NGC 1193 and $f_{\rm 0}=53.597\pm 3.789$ stars arcmin$^{-2}$, $r_{\rm c}=1.190\pm0.056$ arcmin and $f_{\rm bg}=11.318\pm 0.321$ stars arcmin$^{-2}$ for NGC 1798, respectively. At the $r=8$ arcmin limiting radius, the stellar density becomes similar to the background density (a grey horizontal line) as seen in Fig.~\ref{fig:king}a (NGC 1193) and Fig.~\ref{fig:king}b (NGC 1798). Therefore, we concluded that the limiting radii for both clusters are $r_{\rm lim}=8$ arcmin. We considered only the stars inside these limiting radii in further analyses.

\subsection{CMDs and Membership Probabilities of Stars}
\label{section:cmds}

The membership probabilities ($P$) of stars located in each of two cluster regions were calculated applying the Unsupervised Photometric Membership Assignment in Stellar Cluster program \citep[{\sc upmask};][]{Krone-Martins14}. {\sc upmask} uses $k$-means clustering, where $k$ is the number of clusters, to detect spatially concentrated groups and identify the most likely cluster members. As an integer $k$-means is not adjusted directly by the user and the best result from the {\sc upmask} methodology is achieved when the $k$-means value is within 6 to 25 \citep{Krone-Martins14, Cantat-Gaudin20}. We applied {\sc upmask} to calculate stellar membership probabilities by considering each star's five-dimensional astrometric parameters from {\it Gaia} EDR3 \citep{Gaia21}, which contains equatorial coordinates ($\alpha$, $\delta$), proper motion components ($\mu_{\alpha}\cos\delta$, $\mu_{\delta}$), trigonometric parallaxes ($\varpi$), and their uncertainties. During application we scaled these five parameters to unit variance and ran 100 iterations for each clusters to assess cluster membership. The membership probability of a star is defined by the frequency of the group in which it is clustered. We reached the best results when $k$ were set to 12 for NGC 1193 and 15 for NGC 1798. We identified as possible cluster members those stars brighter than $G=20$ mag with membership probabilities $P\geq 0.5$ that we identified as possible members of clusters. This led to 735 possible members for NGC 1193 and 1,536 for NGC 1798. \citet{Cantat-Gaudin20} give the number of stars brighter than $G=18$ mag with the membership probabilities $P > 0.5$ as 215 for NGC 1193 and 218 for NGC 1798. The dissimilarity can be explained by lower precision in the astrometric {\it Gaia} DR2 data compared to {\it Gaia} EDR3, as well as the $G$ magnitude limit of stars used in analyses. With the release of {\it Gaia} EDR3 data, the precision of astrometric and photometric measurements increased with reference to {\it Gaia} EDR2 data. For the {\it Gaia} EDR3 release accuracy of trigonometric parallaxes increased by 30 percent and uncertainties reduced by nearly 40\%, proper motion accuracy increased by a factor of 2 and associated uncertainties improved by a factor $\sim 2.5$. Moreover, the precision of photometric data and celestial positions are better in terms of homogeneity \citep{Gaia21}.

\begin{figure*}
\centering
\includegraphics[scale=0.52, angle=0]{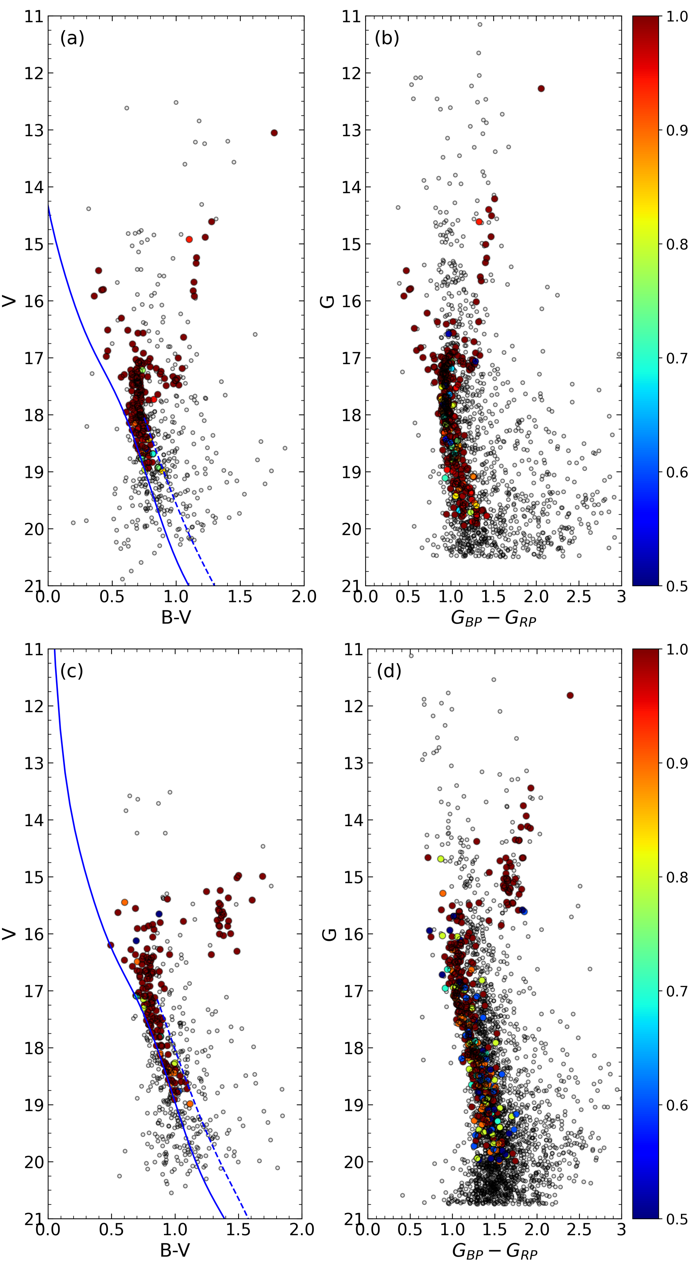}
\caption{$V\times (B-V)$ and $G\times (G_{\rm BP}-G_{\rm RP})$ CMDs of NGC 1193 (a, b) and NGC 1798 (c, d). The blue dot-dashed lines represent the ZAMS \citep{Sung13} including the binary star effect. The membership probabilities of stars that lie within the fitted ZAMS are shown with different colors according to the color scales shown to the right of the figure. These member stars are located within $r_{\rm lim}=8$ arcmin of the cluster centres calculated for NGC 1193 and NGC 1798. Grey dots indicate low probability members ($P<0.5$), or field stars ($P=0$).} 
\label{fig:cmds}
\end {figure*}

\begin{figure}
\centering
\includegraphics[scale=0.28, angle=0]{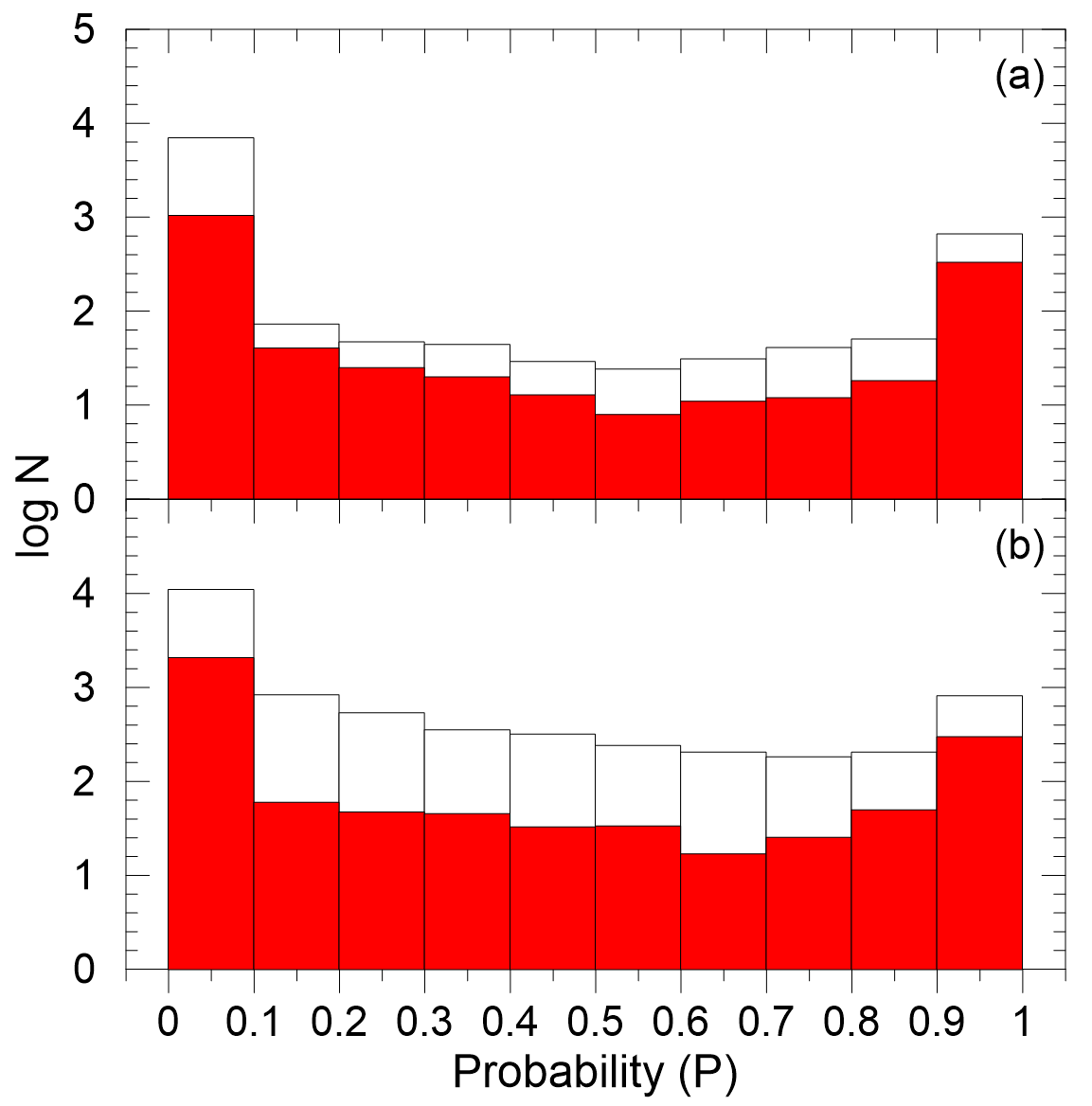}
\caption{Histograms of the membership probabilities for NGC 1193 (a) and NGC 1798 (b). The red colored shading denotes the stars that lie within the main-sequence band and effective cluster radii ($\rm r_{lim} \leq 8'$),  while the white colored bars indicate the membership probabilities of all stars in each cluster's direction.}
\label{fig:prob_hists}
\end {figure}

\begin{figure*}
\centering
\includegraphics[scale=.24, angle=0]{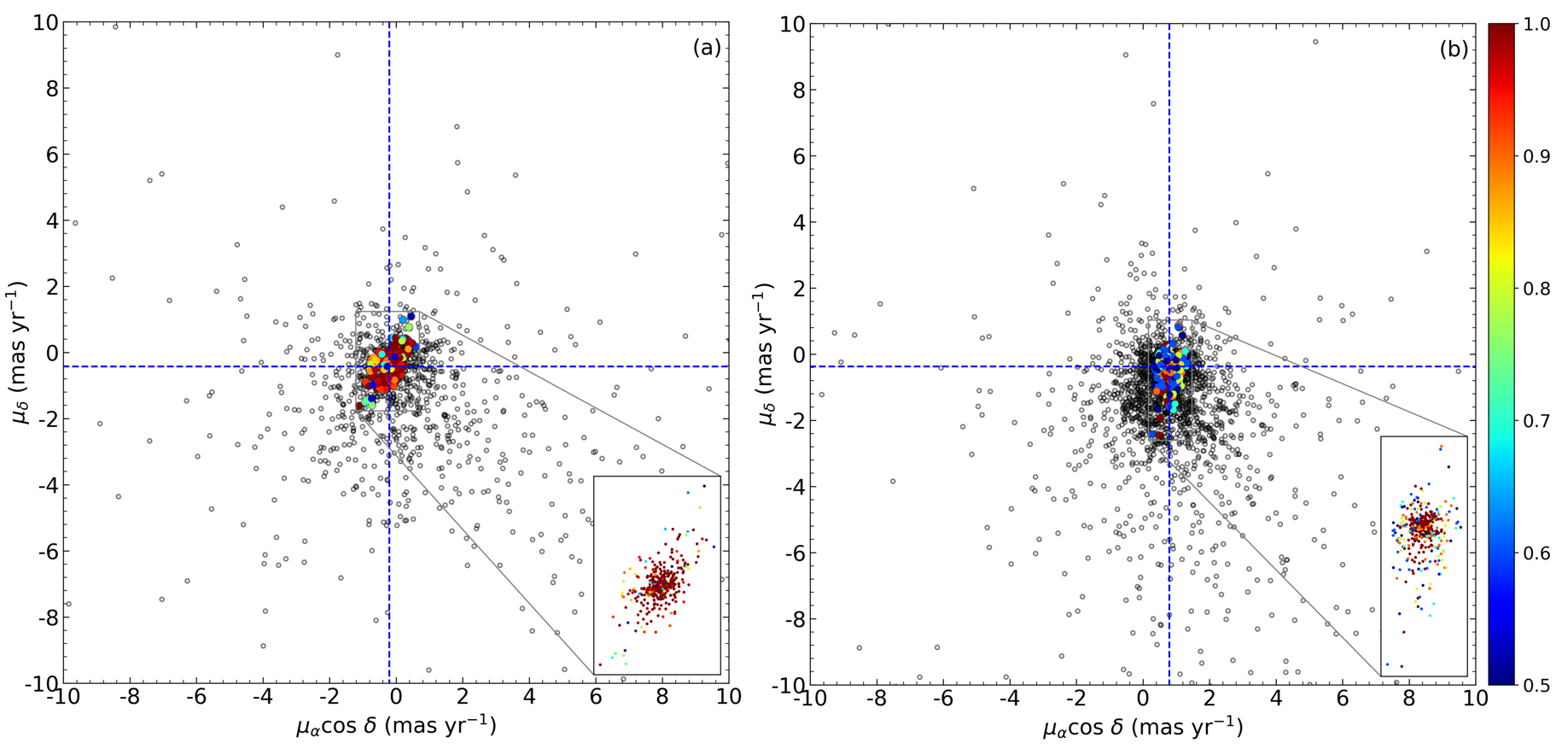}\\
\caption{VPDs of NGC 1193 (a) and NGC 1798 (b). Colored dots identify the membership probabilities of the most likely cluster members according to color scale shown on the right. The zoomed box in panels represents the region of condensation for both clusters in the VPD. Dashed lines are the intersection of the mean proper motion values.}
\label{fig:vpds} 
\end {figure*}

To take into consideration the impact of binary stars in the main-sequences of the studied clusters, we plotted the $V\times (B-V)$ CMDs of the stars within the cluster limiting radii ($r_{\rm lim}$) which we had obtained above for the clusters and then fitted the Zero Age Main-Sequence (ZAMS) of \citet{Sung13} to these diagrams. The ZAMS fitting was by eye according to the stars with the membership probability $P \geq 0.5$ and shifted 0.75 mag towards brighter magnitudes in order to account for the most likely cluster binary stars (\ref{fig:cmds}a and c). During the ZAMS fitting we made sure for each cluster that the main-sequence, turn-off, and giant stars with the membership probabilities $P \geq 0.5$ were selected. The process was resulted in 361 likely member stars for NGC 1193 and 428 for NGC 1798 which lie between the fitted ZAMS curves and are located inside the $r_{\rm lim}$ radii. We used these stars to determine astrophysical parameters of the two clusters. Fig.~\ref{fig:cmds} shows the $V\times (B-V)$ CMDs with the best fitted ZAMS (Figs. \ref{fig:cmds}a and c) and $G\times (G_{\rm BP}-G_{\rm RP}$) CMDs (Figs. \ref{fig:cmds}b and d) with the background and most likely member stars. Fig.~\ref{fig:prob_hists} presents histograms of number of stars located through the two cluster fields versus their membership probabilities. Vector-Point Diagrams (VPDs) were plotted for the stars within the limiting radii and are shown as Fig.~\ref{fig:vpds}. It can be seen from the figure that NGC 1193 (Fig. \ref{fig:vpds}a) and  NGC 1798 (Fig. \ref{fig:vpds}b) are affected by field stars but with the membership selection criteria, the `most likely' cluster stars (shown as the color-scaled points in Fig. \ref{fig:vpds}) can be separated from field stars (grey dots in Fig. \ref{fig:vpds}). The mean proper motion components of the most likely cluster members are ($\mu_{\alpha}\cos \delta$, $\mu_{\delta})=(-0.207 \pm 0.009, -0.431 \pm 0.008$) for NGC 1193 and ($\mu_{\alpha}\cos \delta$, $\mu_{\delta}) = (0.793 \pm 0.006, -0.373 \pm 0.005$) mas yr$^{-1}$ for NGC 1798. Moreover, using these members we obtained mean trigonometric parallaxes of NGC 1193 and NGC 1798 as $\varpi_{\rm Gaia}= 0.191 \pm 0.157$ mas and $\varpi_{\rm Gaia}= 0.203 \pm 0.099$ mas, respectively.

\section{Astrophysical Parameters of the Clusters}
We summarize in this section the processes we performed to determine the astrophysical parameters of NGC 1193 and NGC 1798 \citep[for detailed descriptions on the methodology see][]{Yontan15, Yontan19, Yontan21, Ak16, Bilir06, Bilir10, Bilir16, Bostanci15, Bostanci18, Banks20, Akbulut2021, Koc22}. Color excesses and metallicities of the clusters were derived using two-color diagrams (TCDs), whereas we obtained distance moduli and ages individually by fitting theoretical models on CMDs. 

\subsection{Reddening}

\begin{figure*}
\centering
\includegraphics[scale=0.26, angle=0]{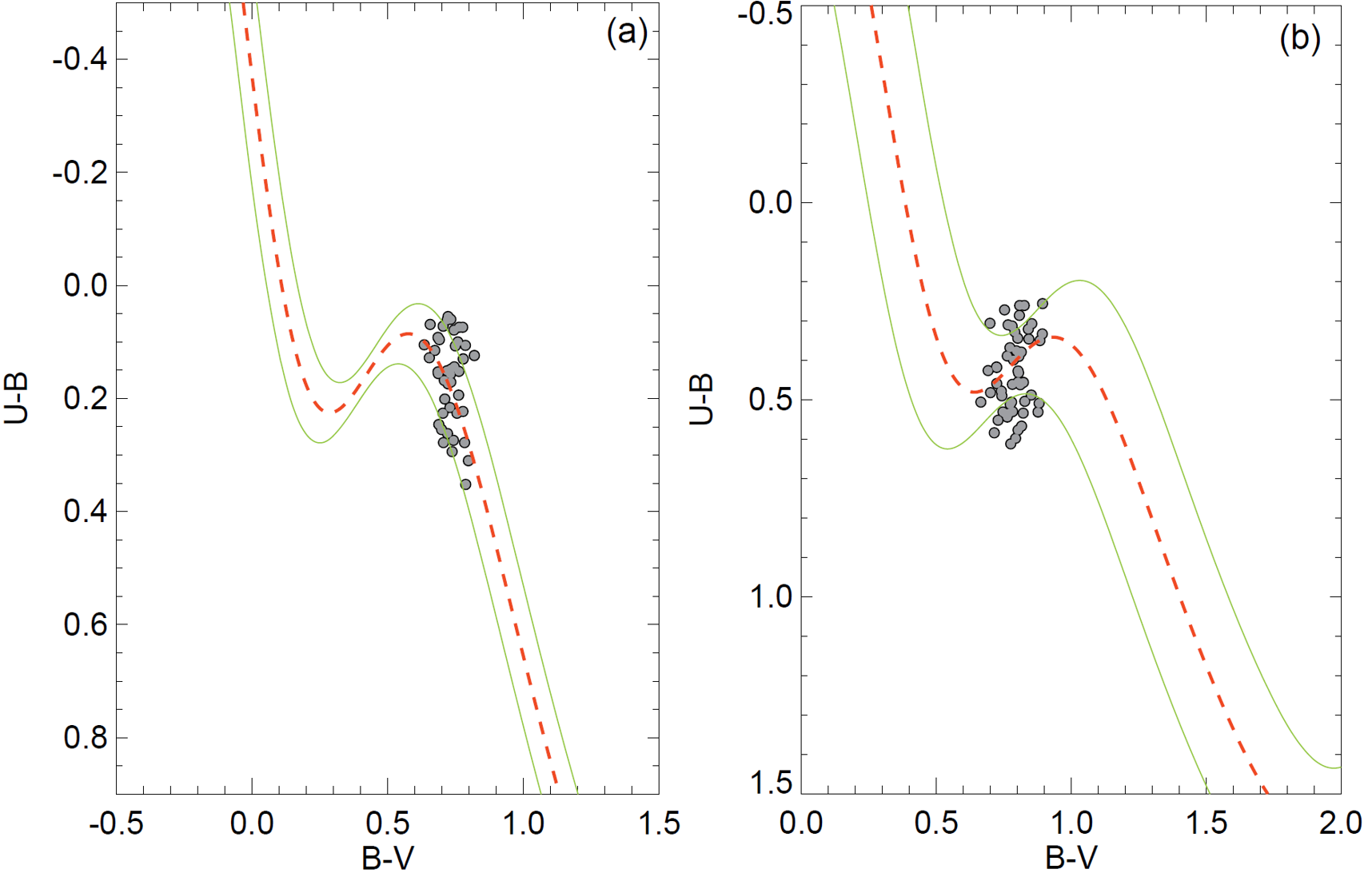}
\caption{Two-color diagrams of the most probable member main-sequence stars in the regions of NGC 1193 (a) and NGC 1798 (b). Red dashed and green solid curves represent the reddened ZAMS given by \citet{Sung13} and $\pm1\sigma$ standard deviations, respectively.
\label{fig:tcds}} 
\end{figure*}

The $E(U-B)$ and $E(B-V)$ color excesses for NGC 1193 and NGC 1798 were derived using $(U-B)\times (B-V)$ TCDs. We selected the main-sequence stars for which simultaneous $U$, $B$, and $V$ magnitudes are available as well as with membership probabilities $P\geq 0.5$. As shown in Fig.~\ref{fig:tcds}, we constructed TCDs for these stars and compared their positions by fitting the solar metallicity de-reddened ZAMS of \citet{Sung13}. The ZAMS was fitted according to the equation $E(U-B)=0.72 \times E(B-V) + 0.05\times E(B-V)^2$ \citep{Garcia88} by applying $\chi^2$ optimisation with steps of 0.001 mag. The best solutions for $E(B-V)$ and $E(U-B)$ values are those corresponding to the minimum $\chi^2$, being $E(B-V)=0.150\pm 0.037$ mag for NGC 1193 and $E(B-V)=0.505\pm 0.100$ mag for NGC 1798. The errors of color excesses are determined as $\pm 1\sigma$ deviations, and are presented as the green lines in Fig.~\ref{fig:tcds}. When we compared the reddening estimated for NGC 1193, we concluded that it is in a good agreement within the errors with the values ($0.10 \leq E(B-V) \leq 0.19$ mag) given by different authors \citep{Kaluzny88, Tadross05, Kyeong08}. For NGC 1798, our finding result is compatible with the values given by \citet[][$E(B-V)=0.51 \pm 0.04$ mag]{Park99} and \citet[][$E(B-V)=0.47 \pm 0.07$ mag]{Oralhan15}.

\begin{figure}[!t]
\centering
\includegraphics[scale=0.35, angle=0]{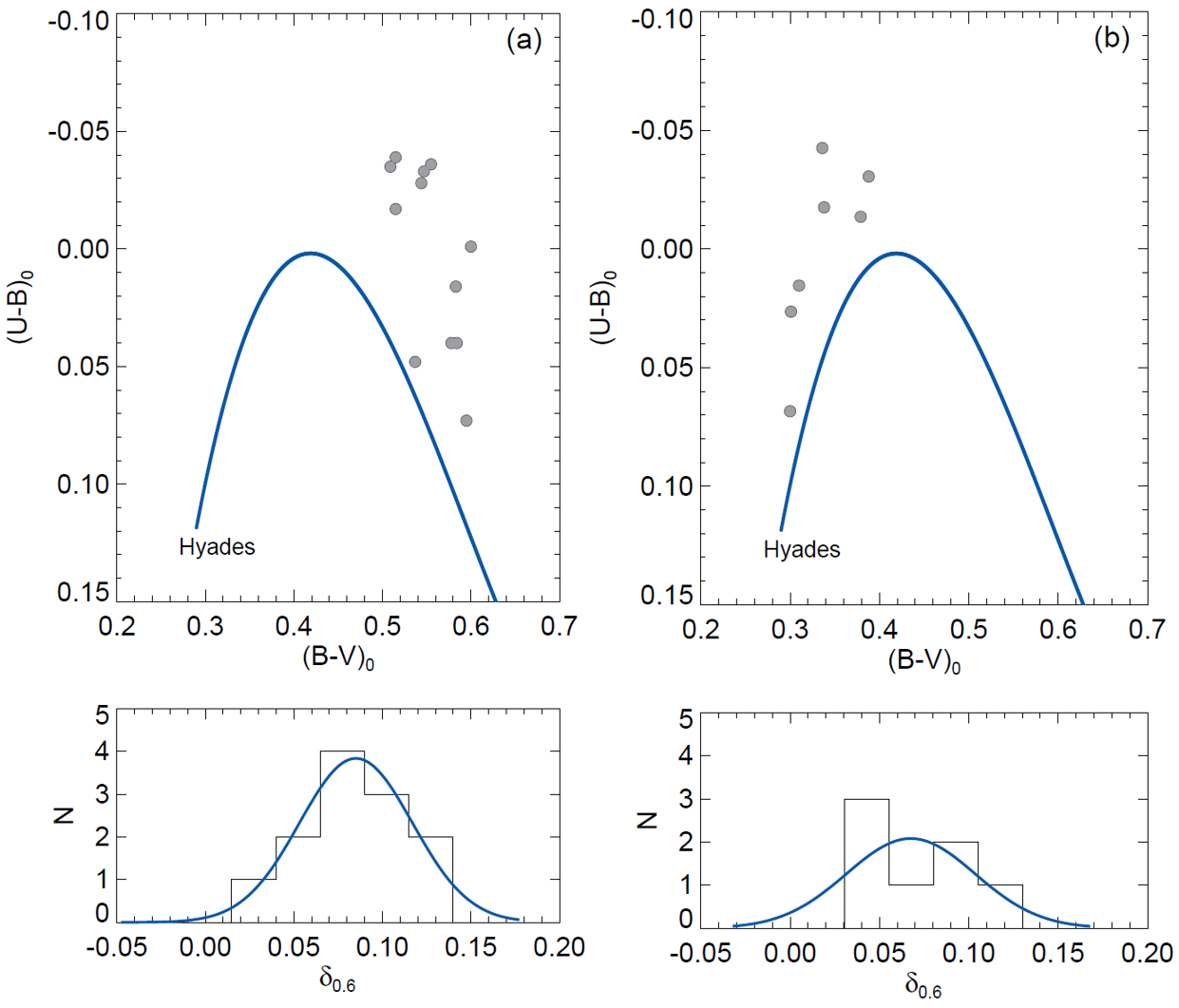}
\caption{Two-color diagrams (upper panels) and the distributions of normalised $\delta_{0.6}$ (lower panels) for NGC 1193 (a) and NGC 1798 (b). The solid blue lines in the upper and lower panels represent the main-sequence of Hyades and Gaussian models which were fitted to the histograms, respectively.
\label{fig:hyades}} 
\end {figure}

\begin{table}[!t]
\setlength{\tabcolsep}{1.5pt}
\renewcommand{\arraystretch}{1.2}
  \centering
  \caption{Metallicities calculated for two clusters. $N$ is the number of member stars used in the analyses.}
    {\small
    \begin{tabular}{lclc|lclc}
    \hline
\multicolumn{4}{c}{NGC 1193} & \multicolumn{4}{c}{NGC 1798}\\
\hline
\multicolumn{1}{l}{$\langle[{\rm  Fe/H}]\rangle$} &  $N$ & \multicolumn{1}{l}{Survey/} & Ref & \multicolumn{1}{l}{$\langle[{\rm  Fe/H}]\rangle$} &  $N$ & Survey/  & Ref \\
\multicolumn{1}{l}{(dex)} & & \multicolumn{1}{l}{Catalog/} & & \multicolumn{1}{l}{(dex)} & & \multicolumn{1}{c}{Catalog/} & \\
\multicolumn{1}{c}{} & & \multicolumn{1}{l}{Telescope} & & \multicolumn{1}{c}{} & & \multicolumn{1}{c}{Telescope} & \\
\hline
$-0.51\pm 0.09$   & 4    & KPNO        & (01) & $-0.18\pm 0.02$    & 4 & KPNO  & (10)\\
$-0.22\pm 0.14$   & 2    & HET         & (02) & $-0.165$           & 4 & KPNO  & (11)\\
$-0.17\pm 0.13$   & 1    & HET         & (03) & $-0.18\pm 0.01$    & 4 & KPNO  & (06)\\
$-0.22\pm 0.01$   & 1    & PASTEL      & (04) & $-0.294$           & 4 & KPNO  & (12)\\
$-0.17$           & 1    & HET         & (05) & $-0.34\pm 0.01$    & 4 & KPNO  & (13)\\
$-0.25\pm 0.01$   & 2    & APOGEE DR14 & (06) & $-0.200\pm 0.006$  & 4 & KPNO  & (14) \\
$-0.34\pm 0.01$   & 3    & APOGEE DR16 & (07) & $-0.30\pm 0.02$    & 4 & KPNO  & (15)\\
$-0.320\pm 0.012$ & 1    & GALAH DR3   & (08) & $-0.27\pm 0.03$    & 4 & KPNO  & (07)\\
$-0.30\pm 0.06$   & 12   & SPMO        & (09) & $-0.267\pm 0.007$  & 4 & KPNO  & (08)\\
                  &      &             &      & $-0.20\pm 0.07$    & 7 & SPMO  & (09)\\
\hline
    \end{tabular} }%
    \\ 
(01) \citet{Friel02}, (02) \citet{Friel10}, (03) \citet{Jacobson13}, (04) \citet{Heiter14}, 
(05) \citet{Overbeek16}, (06) \citet{Carrera19}, (07) \citet{Donor20}, (08) \citet{Spina21}, (09) This study, (10) \citet{Donor18}, (11) \citet{Ting18}, (12) \citet{Ting19}, (13) \citet{Hasselquist20}, (14) \citet{Sit20}, (15) \citet{Olney20}
  \label{tab:metal_abundances}%
\end{table}%

\subsection{Metallicities}

The determination of photometric metallicities of the two clusters employed the method given by \citet{Karaali03a, Karaali03b,Karaali11}. This method is based on F and G type main-sequence stars and their UV-excesses as well as the stars whose color index range correspond to $0.3\leq (B-V)_0\leq0.6$ mag \citep{Eker18, Eker20}. We selected F-G type main-sequence stars within the range $0.3\leq (B-V)_0\leq0.6$ mag after calculating the intrinsic $(B-V)_0$ and $(U-B)_0$ colors of the most likely cluster member ($P\geq 0.5$) stars. To determine the difference between the $(U-B)_0$ color indices of cluster stars and Hyades main sequence which corresponds to the same $(B-V)_0$ color indices, we constructed $(U-B)_0\times(B-V)_0$ TCDs. This difference between cluster and Hyades stars is defined as the UV-excess which is expressed by the equation of $\delta =(U-B)_{\rm 0,H}-(U-B)_{\rm 0,S}$, where H and S denote the Hyades and cluster stars respectively, which implies the same $(B-V)_0$ color indices. By calibrating $(B-V)_0$ of stars to $(B-V)_0 = 0.6$ mag (i.e., $\delta_{0.6}$) we normalised the UV excess and plotted the histogram of normalised $\delta_{0.6}$ values. To calculate the mean $\delta_{0.6}$, we fitted a Gaussian to the distribution. Taking into account the Gaussian peak, the photometric metallicities of the studied clusters are obtained from the equation given by \citet{Karaali11}:
\begin{eqnarray}
{\rm [Fe/H]}=-14.316(1.919)\delta_{0.6}^2-3.557(0.285)\delta_{0.6}+0.105(0.039).
\end{eqnarray}

We identified 12 and 7 F-G type main-sequence stars to calculate the photometric metallicity of NGC 1193 and NGC 1798, respectively. TCDs and the distributions of normalised $\delta_{0.6}$ UV excesses for two clusters are shown in Fig.~\ref{fig:hyades}. The calculated mean $\delta_{0.6}$ values of NGC 1193 and NGC 1798 are 0.085$\pm$0.010 mag and 0.068$\pm$0.011 mag, respectively. The photometric metallicity [Fe/H] value for NGC 1193 is ${\rm [Fe/H]} = -0.30 \pm 0.06$ dex and for NGC 1798 ${\rm [Fe/H]} = -0.20 \pm 0.07$ dex, which correspond to their peak values in the $\delta_{0.6}$ distribution.

The [Fe/H] metallicities were transformed to the mass fraction $Z$ to derive ages of the clusters. For this, the analytic equations of Bovy\footnote{https://github.com/jobovy/isodist/blob/master/isodist/Isochrone.py}$^{,}$ \footnote{The equations are given in lines between 199 and 207 in the code.} for {\sc parsec} \citep{Bressan12} models were used, namely:
\begin{equation}
z_{\rm x}={10^{{\rm [Fe/H]}+\log \left(\frac{z_{\odot}}{1-0.248-2.78\times z_{\odot}}\right)}}
\end{equation}      
and
\begin{equation}
z=\frac{(z_{\rm x}-0.2485\times z_{\rm x})}{(2.78\times z_{\rm x}+1)}.
\end{equation} 
$z$ and $z_{\rm x}$ are the elements heavier than helium and the intermediate operation function, respectively. $z_{\odot}$ is the solar metallicity which was adopted as 0.0152 \citep{Bressan12}. We calculated $z=0.008$ for NGC 1193 and $z=0.010$ for NGC 1798.

Many authors obtained spectroscopic metallicities of NGC 1193 and NGC 1798 based on ground-based observations, as listed in Table~\ref{tab:metal_abundances}. Photometric metallicities calculated in this study are well supported by the spectroscopic studies presented in literature. We conclude that our metallicity findings are reliable. Thus, we adopted our results for the determination of distance moduli and age.

\subsection{Distance Moduli and Age Estimation}

\begin{figure*}[!tb]
\centering
\includegraphics[scale=.4, angle=0]{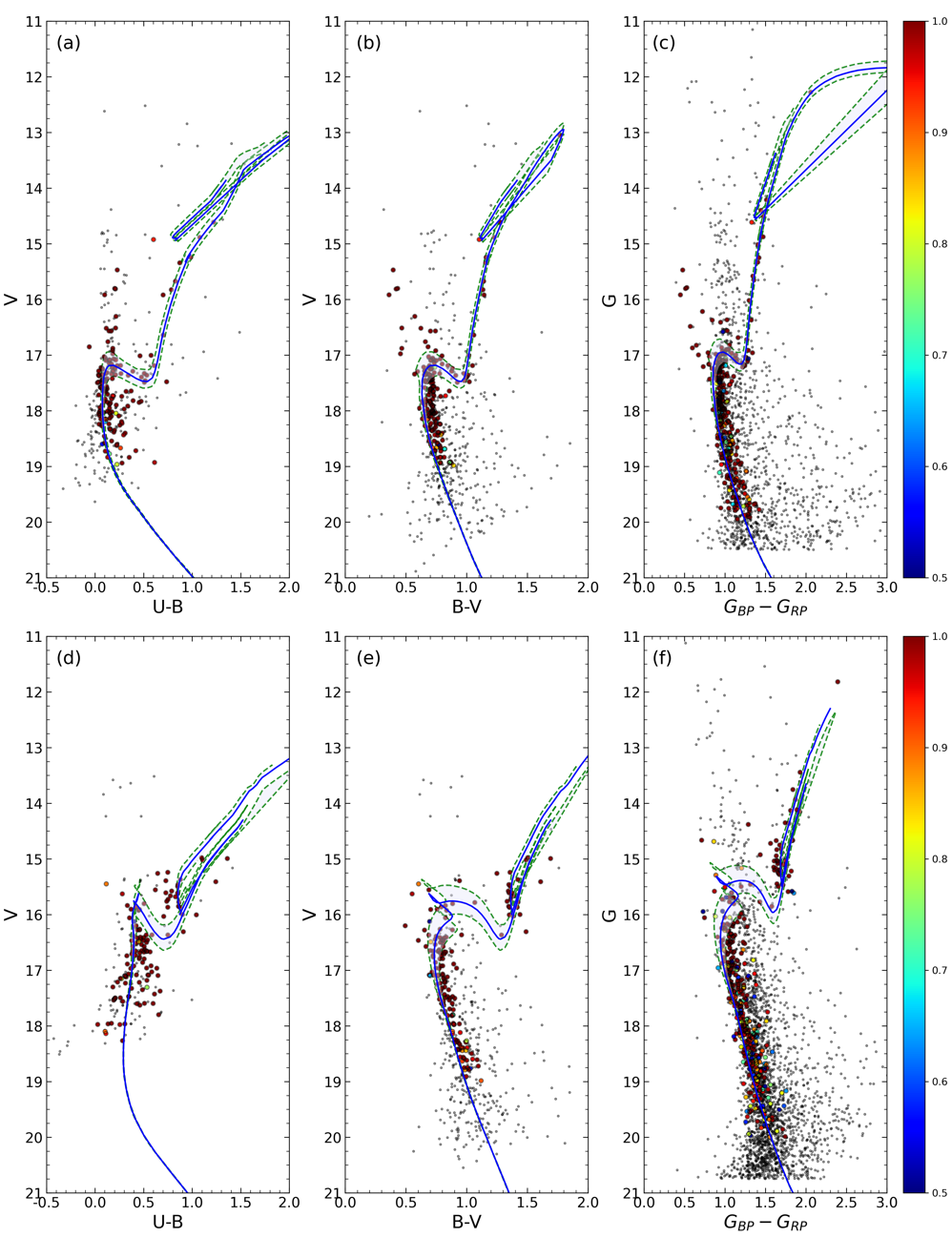}
\caption{CMDs for the NGC 1193 (panels a, b, and c) and NGC 1798 (panels d, e, and f). The differently colored dots represent the membership probabilities according to the color scales shown on the right side of the diagrams. Grey dots indicate low probability members ($P<0.5$), or field stars ($P=0$). The blue lines show the {\sc parsec} isochrones, while the shaded areas surrounding these lines are their associated errors.
\label{fig:age_cmds} }
\end {figure*}

We used {\sc parsec} isochrones \citep{Bressan12}, which contain {\it UBV} filters as well as {\it Gaia} pass-bands, to obtain the distance moduli and ages of the studied clusters simultaneously. To do this, we selected the {\sc parsec} models considering the mass fractions ($z$) estimated for each cluster and compared them to the $V\times (U-B)$, $V\times (B-V)$, and $G\times (G_{\rm BP}-G_{\rm RP})$ CMDs according to member stars ($P\geq 0.5$). Selected isochrones were fitted to CMDs visually by attaching importance to `most likely' member stars which make up the main-sequence, turn-off and giant regions of each cluster. During the fitting process of {\sc parsec} models to the $UBV$ data, we used the $E(B-V)$ values derived above by this study, while for the {\it Gaia} EDR3 data we considered the equation of $E(G_{\rm BP}-G_{\rm RP})= 1.41\times E(B-V)$ \citep{Sun21}. We obtained the error of the distance moduli and distances using the relation given by \citet{Carraro17}. We fitted two more isochrones to estimate age uncertainties considering the spread of the most likely member stars in the turn-off and sub-giant regions of the cluster. The age of such selected isochrones give the higher and lower acceptable values for the estimated cluster ages. The best fit with $z=0.008$ gave the distance moduli and age of NGC 1193 as $\mu=14.191 \pm 0.149$ mag and $t=4.6 \pm 1.0$ Gyr. For NGC 1798, the best fit of $z=0.010$ gave these values as $\mu=14.808 \pm 0.332$ mag and $t=1.3 \pm 0.2$ Gyr, respectively. The distances of the clusters corresponding to the estimated distance moduli are also $d_{\rm iso}=5562 \pm 381$ pc for NGC 1193 and $d_{\rm iso}=4451 \pm 728$ pc for NGC 1798. The $V\times (U-B)$, $V\times (B-V)$, and $G\times (G_{\rm BP}-G_{\rm RP})$ CMDs with the best fit isochrones and associated errors are shown in Fig.~\ref{fig:age_cmds}. 

The isochrone-based distance for NGC 1193 as estimated by this study is compatible with the result given by \citet[][$d=5.25\pm 0.24$ kpc]{Tadross05}. As well, the estimated age of the cluster is in a good agreement with the value of \citet[][$t=5.0\pm 1.3$ Gyr]{Kyeong08}. For NGC 1798, the derived distance matches well within the errors with the result of \citet[][$d=4.2\pm 0.3$ kpc]{Park99}. The age of the cluster is coherent with the findings given by \citet[][$t=1.4\pm 0.2$ Gyr]{Park99} and \citet[][$t=1.6$ Gyr]{Maciejewski07}.

Applying the linear equation of $\varpi \: ({\rm mas})=1000/d \: ({\rm pc})$, we converted iso\-chrone distances to trigonometric parallaxes for the two clusters. This indicated that the parallax distances of NGC 1193 and NGC 1798 are $\varpi_{\rm iso}= 0.180 \pm 0.012$ mas and $\varpi_{\rm iso}= 0.225 \pm 0.037$ mas, respectively. It is concluded that these values are in good agreement with the {\it Gaia} EDR3 trigonometric parallax distances for both clusters.

\section{Kinematics and Galactic Orbit Parameters of Clusters}

The {\sc MWPotential2014} \citep{Bovy15} algorithm,  one of the potential functions defined in {\sc galpy} (the galactic dynamics library \citep{Bovy15}\footnote{See also https://galpy.readthedocs.io/en/v1.5.0/}), was applied to calculate the space velocity components and galactic orbital parameters for NGC 1193 and NGC 1798. The algorithm assumes an axisymmetric potential for the Milky Way galaxy. We adopted the galacto\-cen\-tric distance to be $R_{\rm GC}=8$ kpc, the Solar circular velocity of  $V_{\rm rot}=220$ km s$^{-1}$ \citep{Bovy15, Bovy12}, and the Solar distance from the galactic plane as $27\pm 4$ pc \citep{Chen00}. Since the {\sc MWPotential2014} code comprises bulge, disk, and halo potentials of the Milky Way, we assumed that it well represents the Galaxy.  

\citet{Bovy15} defined the bulge component as a spherical power law density profile, given as follow:

\begin{equation}
\rho (r) = A \left( \frac{r_{\rm 1}}{r} \right) ^{\alpha} \exp \left[-\left(\frac{r}{r_{\rm c}}\right)^2 \right] \label{eq:rho}
\end{equation} 
where $r_{\rm 1}$ is the present reference radius, $r_{\rm c}$ the cut-off radius,  $A$ the amplitude that is applied to the potential in mass density units, and $\alpha$ is the inner power.  We adopted the potential presented by \citet{Miyamoto75} for the galactic disk component:

\begin{equation}
\Phi_{\rm disk} (R_{\rm GC}, Z) = - \frac{G M_{\rm d}}{\sqrt{R_{\rm GC}^2 + \left(a_{\rm d} + \sqrt{Z^2 + b_{\rm d}^2 } \right)^2}} \label{eq:disc}
\end{equation}
$R_{\rm GC}$ is the distance from the galactic centre, $Z$ the vertical distance from the galactic plane, $G$ the universal gravitational constant, $M_{\rm d}$ the mass of the galactic disk, and $a_{\rm d}$ and $b_{\rm d}$ being the scale-length and scale-height of the disk, respectively.

The potential for the halo component was obtained by \citet{Navarro96} as:
\begin{equation}
\Phi _{\rm halo} (r) = - \frac{G M_{\rm s}}{R_{\rm GC}} \ln \left(1+\frac{R_{\rm GC}}{r_{\rm s}}\right) \label{eq:halo}
\end{equation} 
where $M_{\rm s}$ is the mass of the dark matter halo of the Milky Way and $r_{\rm s}$ is its radius.

To determine the spacial velocities and galactic orbit parameters of NGC 1193 and NGC 1798, we used the equatorial coordinates, proper motion components, distances, and radial velocity data with their uncertainties in the calculations.  These values are listed in Table~\ref{tab:Final_table} (on page~\pageref{tab:Final_table}). We performed kinematic and dynamic analyses with 1 Myr steps over a 3.5 Gyr integration time. We considered the proper motion components and distances of the two clusters as derived by this study (see Section~\ref{section:cmds}), while for the radial velocities we used the data of \citet{Donor20} who gave $\langle V_{\rm r}\rangle=-84.7 \pm 0.2 $ km s$^{-1}$ for NGC 1193 and $\langle V_{\rm r}\rangle=2.7 \pm 0.8 $ km s$^{-1}$ for NGC 1798. As a result, we obtained for both clusters estimates of apogalactic distance $R_{\rm a}$, perigalactic  distance $R_{\rm p}$, eccentricity $e$, maximum vertical distance from galactic plane $Z_{\rm max}$, galactic space velocity components ($U$, $V$, $W$), and orbital period $T$. These estimates are listed in Table~\ref{tab:Final_table}. The  space velocity components $(U, V, W)$ were calculated as $(70.95 \pm 0.16, -47.62 \pm 0.10, 5.56 \pm 0.59)$ for NGC 1193 and $(-7.18 \pm 1.50, -14.64 \pm 2.27, 9.13 \pm 1.69)$ for NGC 1798. In their study based on {\it Gaia} DR2 astrometric data \citep{Gaia18}, \citet{Soubiran18} derived the space velocity components for NGC 1193 as $(U, V, W)$ = ($68.84 \pm 0.53$, $-46.77 \pm 0.54$, $9.00 \pm 0.65$) and for NGC 1798 as $(U, V, W)$ = ($7.50 \pm 0.41$, $-16.63 \pm 0.50$, $12.85 \pm 0.39$) km s$^{-1}$. These results are in good agreement with the values calculated in the study. The correction to the local standard of rest (LSR), given by \citet{Coskunoglu11} as ($U, V, W$) = $(8.83 \pm 0.24$, $14.19 \pm 0.34$, $6.57 \pm 0.21)$ km s$^{-1}$, was applied to the space velocity components. The derived LSR corrected space velocity components are $(U, V, W)_{\rm LSR}$ = ($79.78 \pm 0.29$, $-33.43 \pm 0.35$, $12.13 \pm 0.62$) for NGC 1193 and $(U, V, W)_{\rm LSR}$ = ($1.65 \pm 1.51$, $-0.45 \pm 2.30$, $15.70 \pm 1.70$) km s$^{-1}$ for NGC 1798.  Moreover, the space velocities of NGC 1193 were calculated to be $87.35 \pm 0.77$ km s$^{-1}$ and $15.79 \pm 3.23$ km s$^{-1}$ for NGC 1798.

\begin{figure*}[!tb]
\centering
\includegraphics[scale=1.2, angle=0]{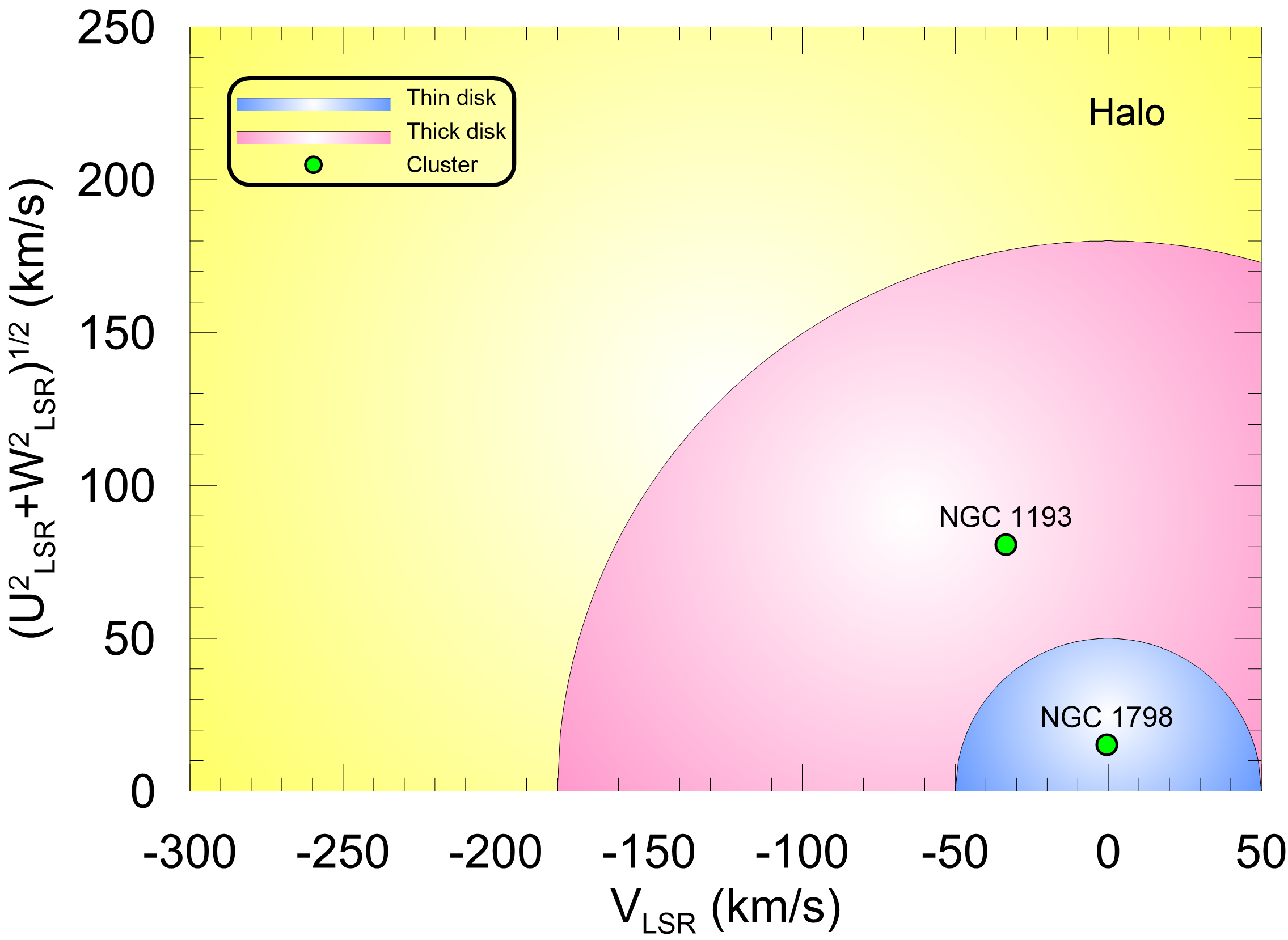}
\caption{Toomre diagram for NGC 1193 and NGC 1798. Blue, pink and yellow regions show thin disk, thick disk and halo populations, respectively.
\label{fig:Toomre_diagram} }
\end {figure*}

Considering the space velocity components of stars in different Galactic populations, \citet{Schuster12} divided the stars into thin disk ($-50<V_{\rm LSR}$ km/s), thick disk ($-180<V_{\rm LSR}\leq -50$ km/s) and halo ($V_{\rm LSR}\leq 180$ km/s) groups. Figure~\ref{fig:Toomre_diagram} shows the positions of the clusters according to \citet{Schuster12}'s kinematic criteria. According to these criteria, the open clusters NGC 1193 and NGC 1798 appear to be members of the thick disk and thin disk populations, respectively. Considering the metal abundance range of NGC 1193 ($-0.51\leq [{\rm Fe/H]}\leq -0.17$ dex, see Table~\ref{tab:metal_abundances}), it is concluded that the cluster belongs to the metal-rich side of the thick-disk population. 

\begin{figure*}
\centering
\includegraphics[width=\textwidth]{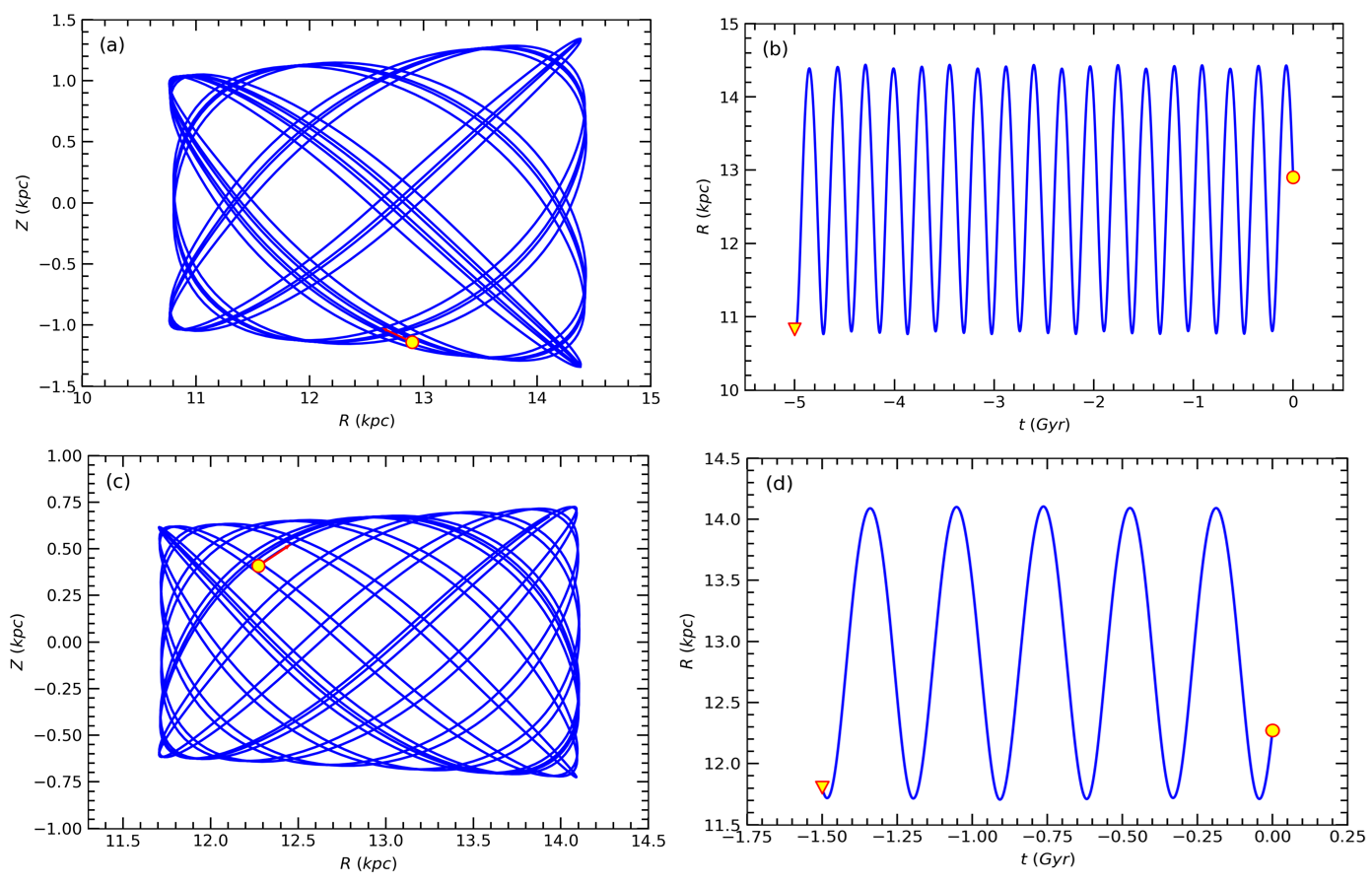}
\caption{\label{fig:galactic_orbits}
The galactic orbits and birth radii of NGC 1193 (a,b) and NGC 1798 (c,d) in the $Z \times R_{\rm GC}$ and $R_{\rm GC} \times t$  planes. The filled yellow circles and triangles show the present day and birth positions, respectively. Red arrows are the motion vectors of OCs.} 
\end {figure*}

Figure~\ref{fig:galactic_orbits} presents the orbits of NGC 1193 (Fig.~\ref{fig:galactic_orbits}a) and NGC 1798 (Fig.~\ref{fig:galactic_orbits}c) as functions of distance from the galactic center and the galactic plane ($Z \times R_{\rm GC}$ and $R_{\rm GC} \times t $). The birth and present-day locations for the two clusters are marked with yellow triangles and circles in sub-figures \ref{fig:galactic_orbits}b and \ref{fig:galactic_orbits}d. Figures \ref{fig:galactic_orbits}a and \ref{fig:galactic_orbits}c show that both of the clusters entirely orbit outside the solar circle. The orbital eccentricities of NGC 1193 and NGC 1798 are smaller than 0.15, thus their orbits are close to circular. The results of orbital integrations imply that NGC 1193 reaches its maximum vertical distance from the galactic plane at $Z_{\rm max} = 1342 \pm 77$ pc with an orbital period $T=370 \pm 12$ Myr, and these values correspond to $Z_{\rm max} = 725\pm 148$ pc and $T = 381 \pm 23$ Myr for NGC 1798. Considering the age values determined in this study for the clusters, we ran the {\sc galpy} program backwards in time and examined the resulting birth--places. The program indicated that the birth-place radial distances are 10.86 kpc and 11.82 kpc for NGC 1193 and NGC 1798, respectively, meaning that the clusters were born in the metal-poor region outside the solar circle.

\section{Luminosity and Present-day Mass Functions}

\begin{figure*}[!t]
\centering
\includegraphics[scale=1, angle=0]{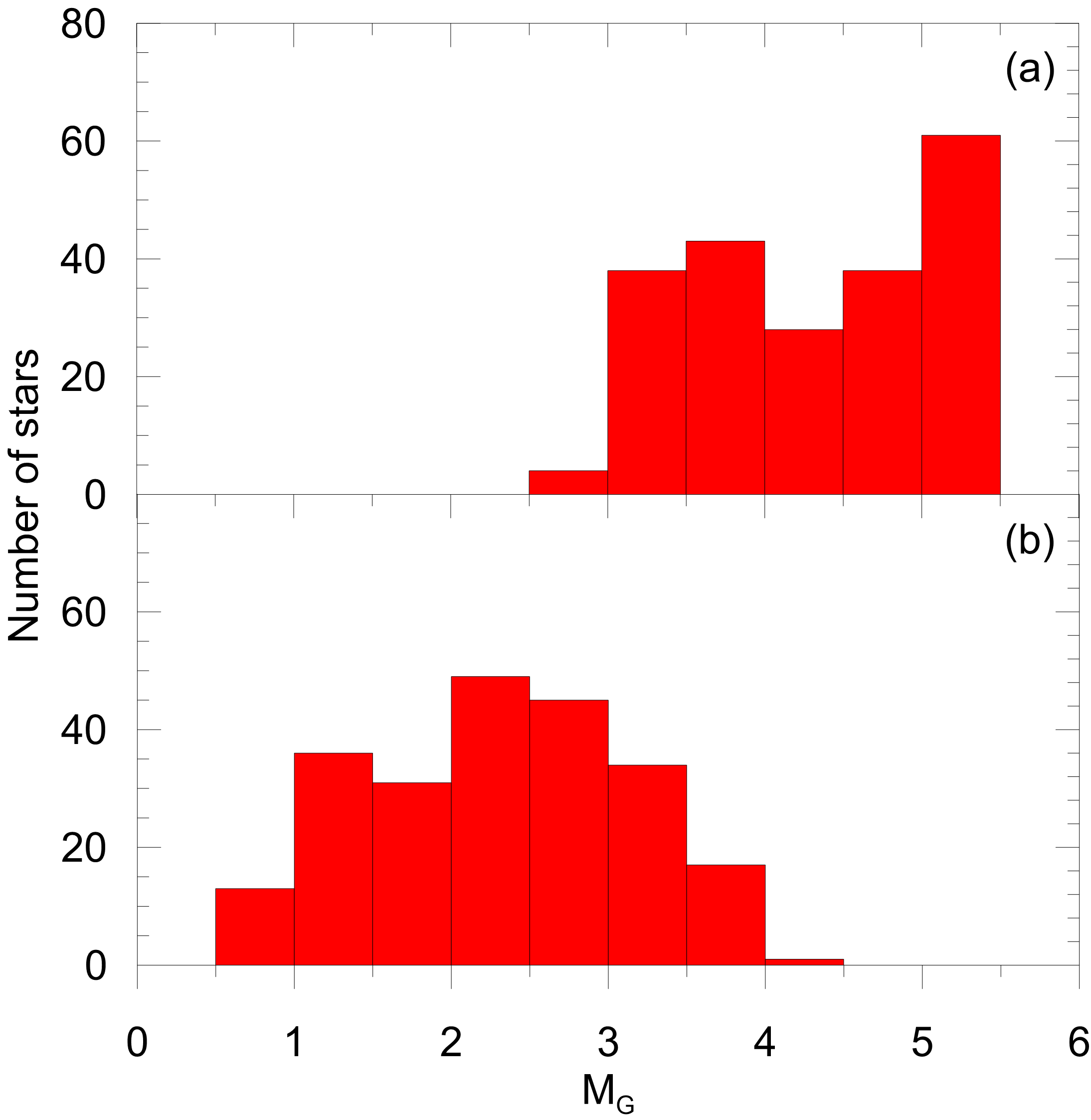}
\caption{\label{fig:luminosity_functions}
The luminosity functions of NGC 1193 (a) and NGC 1798 (b). The histograms show the absolute magnitudes of the main-sequence stars belonging to the clusters.} 
\end {figure*}

The distribution of stars according to their brightness is defined as the luminosity function (LF). We used {\it Gaia} EDR3 photometric data to determine LFs for the two clusters. For this, main-sequence stars located inside the 8 arcmin limiting radii, as derived above, were selected for the two clusters. The magnitude ranges of the chosen stars are within the $17.25\leq G \leq 20$ mag for NGC 1193 and $16.5\leq G \leq 20$ mag for NGC 1798. We converted the $G$ magnitudes of the selected stars to absolute magnitudes with the equation $M_{\rm G} = G-5\times \log d +5+A_{\rm G}$, where $G$ is the apparent magnitude and $d$ the distance derived earlier in this study. $A_{G}$ is extinction for $G$ magnitudes and represented by $A_{G}=0.84\times A_{V}$ \citep{Sun21} (here $A_{V}$ is the extinction for $V$ magnitudes). This led to the absolute magnitude ranges being limited within the $2.5< M_{\rm G}<5.5$ and $0.5< M_{\rm G}<4.5$ mag for NGC 1193 and NGC 1798, respectively. We constructed LF histograms with the step-size 0.5 mag, as shown as Fig.~\ref{fig:luminosity_functions} for both clusters. 

\begin{figure}[!t]
\centering
\includegraphics[scale=1, angle=0]{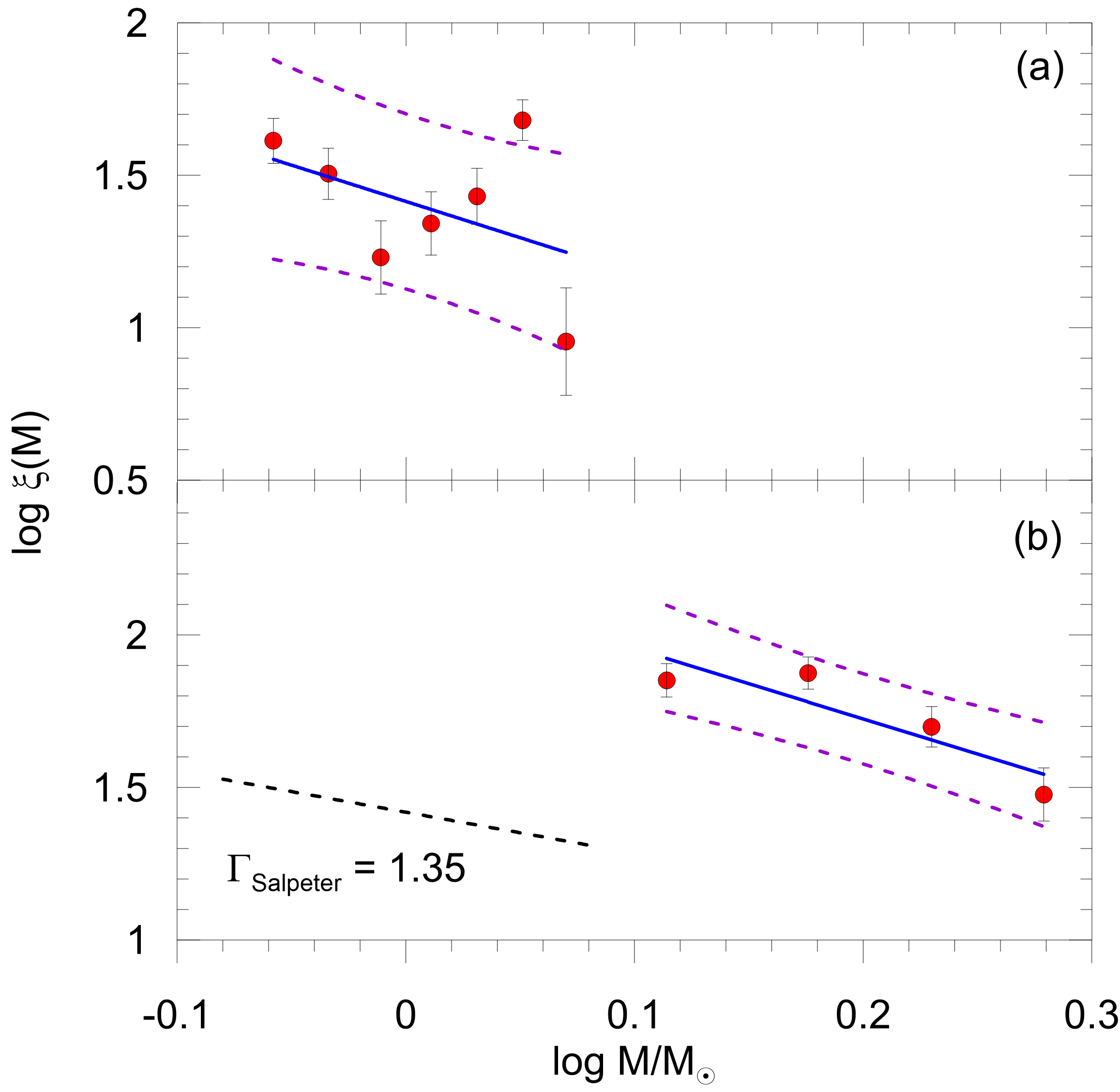}
\caption{\label{fig:mass_functions}
Present-day mass functions of NGC 1193 (a) and NGC 1798 (b) derived from all sample (red circle). The blue and dashed lines represent the mass functions of the open clusters and \citet{Salpeter55}'s mass function, respectively. The purple dashed lines show $\pm1\sigma$ prediction levels.} 
\end {figure}

To convert these LFs to present day mass functions (PDMFs) we employed the {\sc parsec} isochrones \citep{Bressan12}, which gives the ages and metal abundances ($z$) of the clusters. We utilized a high degree polynomial equation between $G$-band absolute magnitudes and masses of theoretical main-sequence stars. The resulting absolute magnitude-mass relation was used to transform the observational absolute $G$ band magnitudes to masses. The number, mass range, and mean mass of main-sequence stars that transformed are 212, $0.85\leq M/ M_{\odot}\leq 1.2$, and 0.99$M_{\rm \odot}$ for NGC 1193, and 226, $1.1\leq M/ M_{\odot}\leq 2$, and 1.53 $M_{\rm \odot}$ for NGC 1798. The mass function PDMF can be approximated by a power law defined as by \citet{Salpeter55}: 

\begin{equation}
{\log(\frac{dN}{dM})=-(1+\Gamma)\times \log(M)+{\rm constant}}
\end{equation} 
Here $dN$ is the number of stars in a mass bin of width $dM$ with a central mass $M$ and $\Gamma$ being the slope of the PDMF. We estimated the slope of the PDMF in both clusters for apparent $G\leq 20$ mag, which corresponds to stars more massive than $0.85 \: M_{\odot}$ in NGC 1193 and $1.1 \:M_{\odot}$ for NGC 1798. The resulting PDMFs with the best fits are presented in Fig.~\ref{fig:mass_functions}. We calculated the slope values to be $\Gamma = 1.38 \pm 2.16$  for NGC 1193 and as $\Gamma = 1.30 \pm 0.21$ for NGC 1798. Since the NGC 1193 cluster is about 5.5 kpc from the Sun, the magnitudes of the main-sequence stars are within a narrow range. This causes the mass range of the main-sequence stars to be limited and the distribution of the mass function to show a large scatter. While the PDMF of the NGC 1193 is compatible with \citet{Salpeter55}'s result of $\Gamma=1.35$, the error of the PDMF is large. This situation is not similar for NGC 1798. Considering the value and error of the PDMF for NGC 1798, it is in agreement with Salpeter's result.

\section{Summary and Conclusion}

We performed photometric, astrometric, and kinematic studies of two old age open clusters, NGC 1193 and NGC 1798, using CCD {\it UBV} and {\it Gaia} EDR3 data. We examined the cluster structure, obtaining basic astrophysical parameters as well as properties of galactic orbits for two clusters. Outcomes of the study are listed in Table~\ref{tab:Final_table} (on page~\pageref{tab:Final_table}) and summarised as follows:

\begin{enumerate}

\item{Performing the RDP analyses, we determined the limiting radii $r_{\rm lim}=8$ arcmin for both clusters. This value corresponds to limiting radii of 12.95 pc and 10.36 pc for NGC 1193 and NGC 1798, respectively. We considered the stars within these limiting radii as potential cluster members and restricted subsequent analysis to this set of stars.
\label{step:rdf}}

\item{The calculation of membership probabilities of stars was made using the {\sc upmask} program together with a five-dimensional parameter space containing the stars' proper motion components, trigonometric parallaxes, and their uncertainties. We considered the stars with probabilities $P\geq0.5$ to be cluster members. Additionally we adopted two more criteria to to clarify cluster membership: 

\begin{table}
  \centering
  \caption{Fundamental parameters of NGC 1193 and NGC 1798.}
  {\small
        \begin{tabular}{lrr}
\hline
Parameter & NGC 1193 & NGC 1798 \\
\hline
($\alpha,~\delta)_{\rm J2000}$ (Sexagesimal) & 03:05:56.64, $+$44:22:58.80 & 05:11:39.36, $+47$:41:27.60\\
($l, b)_{\rm J2000}$ (Decimal)              & 148.8143, $-12.1624$      & 160.7043, $+04$.8500   \\
$f_{0}$ (stars arcmin$^{-2}$)               & $166.865 \pm 1.573$       & $53.597 \pm 3.789$     \\
$r_{\rm c}$ (arcmin)                        & $0.526 \pm 0.009$         & $1.190 \pm 0.057$      \\
$f_{\rm bg}$ (stars arcmin$^{-2}$)          & $5.225 \pm 0.124$         & $11.318 \pm 0.321$     \\
$r_{\rm lim}$ (arcmin)                      &  8                        & 8                      \\
$r$ (pc)                                    & 12.95                     & 10.36                  \\
$\mu_{\alpha}\cos \delta$ (mas yr$^{-1}$)   & $-0.207 \pm 0.009$        & $0.793 \pm 0.006$      \\
$\mu_{\delta}$ (mas yr$^{-1}$)              & $-0.431 \pm 0.008$        &$-0.373 \pm 0.005$      \\
Cluster members ($P\geq0.5$)                & 361                       & 428                    \\
$\varpi$ (mas)                              & $0.191 \pm 0.157$         & $0.203 \pm 0.099$      \\
$E(B-V)$ (mag)                              & $0.150 \pm 0.037$         & $0.505 \pm 0.100$      \\
$E(U-B)$ (mag)                              & $0.109 \pm 0.027$         & $0.376 \pm 0.073$      \\
$A_{\rm V}$ (mag)                           & $0.465 \pm 0.084$         & $1.566 \pm 0.310$      \\
$[{\rm Fe/H}]$ (dex)                        & $-0.30 \pm 0.06$          & $-0.20 \pm 0.07$       \\
Age (Gyr)                                   & $4.6 \pm 1.0$             & $1.3 \pm 0.2$          \\
Distance modulus (mag)                      & $14.191 \pm 0.149$        & $14.808 \pm 0.332$     \\
Isochrone distance (pc)                     & $5562 \pm 381$            & $4451 \pm 728$         \\
$(X, Y, Z)_{\odot}$ (pc)                    & ($-4651$, 2815, 1172)     & ($-4186$, 1466, 376)   \\
$R_{\rm GC}$ (kpc)                          & 12.96                     & 12.27                  \\
PDMF slope                                  &$-1.38\pm 2.16$            & $-1.30 \pm 0.21$       \\
$U_{\rm LSR}$ (km/s)                        & $79.78 \pm 0.29$          & $1.65 \pm 1.51$        \\
$V_{\rm LSR}$ (km/s)                        & $-33.43 \pm 0.35$         & $-0.45 \pm 2.30$       \\
$W_{\rm LSR}$ (km/s)                        & $12.13 \pm 0.62$          & $15.70 \pm 1.70$       \\
$S_{_{\rm LSR}}$ (km/s)                     & $87.35 \pm 0.77$          & $15.79 \pm 3.23$       \\
$R_{\rm a}$ (kpc)                           & $14.44 \pm 0.34$          & $14.11 \pm 0.30$       \\
$R_{\rm p}$ (kpc)                           & $10.80 \pm 0.43$          & $11.72 \pm 0.50$       \\
$z_{\rm max}$ (pc)                          & $1342 \pm 77$             & $725 \pm 148$          \\
$e$                                         & $0.144 \pm 0.008$         & $0.092 \pm 0.011$      \\
$P$ (Myr)                                   & $370 \pm 12$              & $381 \pm 23$           \\
Birthplace (kpc)                            & 10.86                     & 11.82                  \\
\hline
        \end{tabular}%
    } 
    \label{tab:Final_table}%
\end{table}%

\begin{enumerate}
    \item {binary star contamination in the cluster main-sequences which was interpreted by the de-reddened ZAMS fitted to $V\times (B-V)$ CMDs with a shift of $+0.75$ mag in the $V$ band, and}
    \item{within the limiting radii determined in the study (as per step~\ref{step:rdf}).}
\end{enumerate}
Consequently, we selected the stars inside the clusters' limiting radii, within best-fitting ZAMS and with the membership probability $P\geq0.5$ as `real' members of two clusters. Thus we identified 361 and 428 stars as most likely members of NGC 1193 and NGC 1798, respectively.}

\item{The reddening and photometric metallicities of the two clusters were derived separately using CCD {\it UBV} \hspace{0.5pt} TCDs. The reddening analyses were performed by fitting de-reddened ZAMS to main sequence member stars. Photometric metallicity was based on the comparison of F-G type main sequence members with the Hyades main-sequence. The reddening and photometric metallicity for NGC 1193 are $E(B-V)=0.150 \pm 0.037$ mag and [Fe/H]=$-0.30 \pm 0.06$ dex, respectively. The corresponding values for NGC 1798 are $E(B-V)=0.505 \pm 0.100$ mag and [Fe/H]=$-0.20 \pm 0.07$ dex.}

\item{The distance moduli, distance, and age of the NGC 1193 were derived as $\mu_{\rm V}=14.191\pm 0.149$ mag, $d=5562\pm 381$ pc, and $t=4.6\pm 1$ Gyr, respectively. Similarly $\mu_{\rm V}=14.808\pm 0.332$ mag, $d=4451\pm728$ pc, and $t=1.3\pm 0.2$ Gyr were calculated for NGC 1798. These results were obtained by simultaneously fitting {\sc parsec} isochrones on the {\it UBV} and {\it Gaia} EDR3 photometric CMDs utilizing the  most likely member stars according to reddening and metallicities derived in the study.}

\item{Mean proper motion components were calculated as ($\mu_{\alpha}\cos \delta, \mu_{\delta}) = (-0.207 \pm 0.009, -0.431 \pm 0.008$) mas yr$^{-1}$ for NGC 1193 and ($\mu_{\alpha}\cos \delta, \mu_{\delta})\\ = (0.793 \pm 0.006, -0.373 \pm 0.005$) mas yr$^{-1}$ for NGC 1798.}  

\item{We estimated mean trigonometric parallaxes using {\it Gaia} EDR3 data of most likely members for two clusters. The results are $\varpi_{\rm Gaia}= 0.191 \pm 0.157$ mas for NGC 1193  and $\varpi_{\rm Gaia}= 0.203 \pm 0.099$ mas for NGC 1798. We also converted isochrones distances to trigonometric parallaxes by applying the linear equation $\varpi \: ({\rm mas})=1000/d \: ({\rm pc})$ and found $\varpi_{\rm iso}= 0.180 \pm 0.012$ mas for NGC 1193  and $\varpi_{\rm iso}= 0.225 \pm 0.037$ mas for NGC 1798. For both clusters our derived trigonometric parallaxes values calculated from isochrone fitting distances are well supported by the values determined from {\it Gaia} EDR3 trigonometric parallaxes of member stars.}

\item{Space velocities and galactic orbital parameters show that NGC 1193 belongs to the thick-disk population, whereas NGC 1798 is a member of the thin-disk population. Moreover, both of the clusters orbit completely outside the solar circle.}

\item{We found NGC 1193 and NGC 1798 were born outside the solar circle with the birth radii of 10.86 and 11.82 kpc from the Galactic centre, respectively. These birth radii indicate the metal-poor formation region and support the metallicities calculated in the study for the two clusters.}

\item{Present day mass function slopes of $\Gamma=1.38\pm 2.16$ and $\Gamma=1.30 \pm 0.21$ were derived for NGC 1193 and NGC 1798, respectively. While the results for two clusters are in good agreement with the value of \citet{Salpeter55}, that for NGC 1193 possesses a large uncertainty. We concluded that because of its distance, the main-sequence stars of NGC 1193 are limited within narrow magnitudes.}

\end{enumerate}


\section*{Acknowledgments}
The observations of this publication were made at the National Astronomical Observatory, San Pedro M{\'a}rtir, Baja California, M{\'e}xico, and the authors wish to thank the staff of the Observatory for their assistance during these observations. This research has made use of the WEBDA database, operated at the Department of Theoretical Physics and Astrophysics of the Masaryk University. This research has made use of NASA's Astrophysics Data System. We also made use of VizieR and Simbad databases at CDS, Strasbourg, France as well as data from the European Space Agency (ESA) mission \emph{Gaia}\footnote{https://www.cosmos.esa.int/gaia}, processed by the \emph{Gaia} Data Processing and Analysis Consortium (DPAC)\footnote{https://www.cosmos.esa.int/web/gaia/dpac/consortium}. Funding for DPAC has been provided by national institutions, in particular the institutions participating in the \emph{Gaia} Multilateral Agreement. IRAF was distributed by the National Optical Astronomy Observatory, which was operated by the Association of Universities for Research in Astronomy (AURA) under a cooperative agreement with the National Science Foundation. PyRAF is a product of the Space Telescope Science Institute, which is operated by AURA for NASA. We thank the University of Queensland for collaboration software.

\end{document}